\documentclass[prd,twocolumn,amsmath,nofootinbib]{revtex4}
\usepackage{url}
\usepackage{aas_macros,psfig,amsmath}
\begin{document}

\def\bi#1{\hbox{\boldmath{$#1$}}}
\def\sun{\hbox{$\odot$}}
\def\farcs{\hbox{$.\!\!^{\prime\prime}$}}

\newcommand{\be}{\begin{equation}}
\newcommand{\ee}{\end{equation}}
\newcommand{\bea}{\begin{eqnarray}}
\newcommand{\eea}{\end{eqnarray}}

\newcommand{\band}[2]{{^{#1}\!{#2}}}
\newcommand{\lexp}{\mathop{\langle}}
\newcommand{\rexp}{\mathop{\rangle}}
\newcommand{\rexpc}{\mathop{\rangle_c}}

\def\bi#1{\hbox{\boldmath{$#1$}}}

\def\affilmrk#1{$^{#1}$}
\def\affilmk#1#2{$^{#1}$#2;}

\def\ptonp{1}
\def\penn{2}
\def\mit{3}
\def\pton{4}
\def\apo{5}
\def\nyu{2}
\def\drexel{4}
\def\osu{5}
\def\fnal{6}
\def\chicago{7}
\def\hopkins{8}
\def\pitt{9}
\def\tucson{10}
\def\colorado{11}
\def\cmu{12}
\def\hawaii{13}
\def\barcelona{16}
\def\sussex{17}
\def\tokyo{18}
\def\flagstaff{19}
\def\michigan{20}
\def\rochester{21}
\def\psu{22}
\def\efi{23}

\title{SDSS galaxy bias from halo mass-bias 
relation and its cosmological implications}

\author{
Uro\v s Seljak\affilmrk{\ptonp},
Alexey Makarov\affilmrk{\ptonp},
Rachel Mandelbaum\affilmrk{\ptonp},
Christopher M. Hirata\affilmrk{\ptonp},
Nikhil Padmanabhan\affilmrk{\ptonp},
Patrick McDonald\affilmrk{\ptonp},
Michael R. Blanton\affilmrk{\nyu},
Max Tegmark\affilmrk{\penn,\mit},
Neta A. Bahcall \affilmrk{\pton},
J. Brinkmann \affilmrk{\apo}
}

\address{
\parshape 1 -3cm 24cm
\affilmk{\ptonp}{Physics Department, Princeton University, Princeton, NJ 08544,
USA} \\
\affilmk{\nyu}{Center for Cosmology and Particle Physics,
Department of Physics, New York University, 4 Washington
Place, New York, NY 10003}\\
\affilmk{\penn}{Department of Physics, University of Pennsylvania,
Philadelphia, PA 19101, USA} \\
\affilmk{\mit}{Dept. of Physics, Massachusetts Institute of Technology,
Cambridge, MA 02139} \\
\affilmk{\pton}{Princeton University Observatory, Princeton, NJ 08544,
USA} \\
\affilmk{\apo}{Apache Point Observatory, 2001 Apache Point Rd,
Sunspot, NM 88349-0059, USA}
}

\date{Submitted December 2003.}

\begin{abstract}
We combine the measurements of luminosity dependence of bias
with the luminosity dependent 
weak lensing analysis of dark matter around galaxies 
to derive 
the galaxy bias and constrain amplitude of mass fluctuations. 
We take advantage of theoretical and simulation 
predictions that predict that while halo bias is rapidly increasing 
with mass for high masses, it is nearly constant in low mass halos.
We use a new weak lensing 
analysis around the same SDSS galaxies to determine their
halo mass probability distribution. 
We use these halo mass probability distributions 
to predict the bias for each luminosity subsample. 
Galaxies below $L_*$ are antibiased with $b<1$ and for these galaxies bias 
is only weakly dependent on luminosity. In contrast, for galaxies above $L_*$
bias is rapidly increasing with luminosity. These observations are
in an excellent agreement with 
theoretical predictions based on weak lensing halo mass determination 
combined with halo bias-mass relations.  
We find that for standard cosmological parameters 
theoretical predictions are able to explain the observed luminosity 
dependence of bias over 6 magnitudes in absolute luminosity. 
We combine 
the bias constraints with those from the WMAP and the SDSS power spectrum analysis
to derive new constraints on bias and $\sigma_8$. 
For the most general parameter space that includes running and neutrino mass
we find $\sigma_8=0.88\pm 0.06$ and $b_*=0.99\pm 0.07$. 
In the context of spatially flat models we improve the limit on the
neutrino mass for the case of 3 degenerate families
from $m_{\nu}<0.6$eV without bias
to $m_{\nu}<0.18$eV with bias (95\% c.l.), which is weakened to 
$m_{\nu}<0.24$eV if running is allowed.
The corresponding limit for 3 massless +
1 massive neutrino is 1.37eV.

\end{abstract}




\maketitle

\setcounter{footnote}{0}

\section{Introduction}

Galaxy clustering has long been recognized as a powerful tool to 
constrain cosmology. Galaxies are assumed to 
trace dark matter on large scales and 
so the galaxy power spectrum can be related to that of 
the
dark matter. The latter depends on several cosmological 
parameters, such as the epoch of matter-radiation equality, baryon 
to dark matter ratio and the primordial power spectrum shape and 
amplitude. 
The key assumption underlying this approach is that galaxies trace 
dark matter up to an overall factor, called the linear bias $b$, so that 
the galaxy and matter power spectra are  related as 
$P_{\delta_{\rm g}}(k)=b^2P_{\delta_{\rm dm}}(k)$, 
where $\delta_{\rm g}$ and $\delta_{\rm dm}$ 
are the galaxy and dark matter density fluctuations, respectively, 
and $P(k)$ is their power spectrum. 

The linear bias
assumption is thought to be accurate on large scales, but becomes 
less and less accurate on small scales, where details of galaxy 
formation play an important role. The exact transition scale 
between the linear and nonlinear regimes does not have to equal that
of dark matter and may
depend on the type of galaxies one is observing, the treatment  
of redshift space distortions
and the cosmological model. For normal galaxies it
is believed to be somewhere 
around $k=0.1-0.2h/{\rm Mpc}$ 
\citep{1998MNRAS.296...10H,2001ApJ...546...20S,2001MNRAS.325.1359S}. 

While the shape of the galaxy power spectrum for $k<0.2h/{\rm Mpc}$ 
has been used to constrain the 
cosmological parameters \citep{2002MNRAS.337.1068P,2003astro.ph.10723T}, 
the overall amplitude is often ignored.
The reason for this is that the galaxies can 
be biased relative to the dark matter and the bias parameter $b$
depends on 
the galaxy properties, such as luminosity or type. 
This has long been observed as a function of 
morphological type \citep{1976ApJ...208...13D}. More recent surveys
emphasized the luminosity dependence of bias, finding that brighter galaxies
cluster more strongly both in 2dF
\citep{2001MNRAS.328...64N} and in SDSS \citep{2002ApJ...571..172Z}.  
These early studies focused on the nonlinear or 
quasi-linear scales below 10$h^{-1}$Mpc, so
it was not clear that their conclusions applied to 
the linear regime. In particular, on small scales 
the clustering strength 
can increase as a function of luminosity 
if the brighter galaxies preferentially populate larger halos with 
many galaxies inside, such as 
groups and clusters. In this case the increase in clustering amplitude
on small scales  is a reflection of the enhanced 
correlations inside the halo 
and is not necessarily a reflection of these 
halos being more correlated among themselves. 
In contrast with these previous studies a recent 
analysis of SDSS galaxy survey data has
focused the analysis on $k<0.2h$/Mpc (scales above 10$h^{-1}$Mpc) and thus 
measured the linear bias directly \citep{2003astro.ph.10725T}. 
This analysis also found that bias increases as a function of luminosity, 
in a similar way as in the previous studies \citep{2001MNRAS.328...64N,2002ApJ...571..172Z}. 
The relative amplitudes of fluctuation power spectra  for different 
luminosity subsamples was found to differ by almost a 
factor of 2.5 from the bright to the faint end. 
It appears therefore that the 
evidence for 
linear bias increasing with luminosity has finally been established.
In this situation
it is not clear which of the galaxies are 
unbiased ($b=1$). In the absence of additional 
information the overall amplitude of galaxy fluctuations thus cannot be 
directly related to that of dark matter. 

There are at least three
ways that have been proposed so far to break the degeneracy. One is to use 
redshift space distortions, which on large scales depend on the 
parameter $\beta=\Omega_m^{0.6}/b$, where $\Omega_m$ is the matter 
density of the universe. Unfortunately, even with modern 
surveys such as 2dF and SDSS, this parameter has considerable statistical 
uncertainty, 
so in itself this method cannot give a sufficiently precise
bias determination. For example, the SDSS analysis gives $\beta=0.5\pm 0.2$
on scales where nonlinear redshift space distortion modeling is reliable 
\citep{2003astro.ph.10725T}.

A second approach is to determine the bias 
from the bispectrum, as was done for 2dF galaxies \citep{2002MNRAS.335..432V}. 
It is difficult to measure the bispectrum
on very large scales, where deviations from nongaussianity are small. 
In \cite{2002MNRAS.335..432V} 
most of the weight comes from  scales with $ 0.2h/{\rm Mpc}<k<0.3 h/$Mpc, 
in which case however it is not obvious that the bias measured there
also applies to the larger scales where the power spectrum is 
measured, given how rapidly the nonlinear effects become 
important. In principle one could verify this with simulations, but
pure N-body 
simulations (used so far) are not sufficient to verify this hypothesis, 
since galaxies do not trace dark matter on small scales and the 
details of how galaxies populate halos change the 
two and three point functions and
depend on the specifics 
of the particular galaxy formation model. 

A third approach is to compare a weak lensing power spectrum
determination, tracing dark matter,
to the galaxy power spectrum or the weak lensing-galaxy cross-correlation. 
This approach can also give the bias directly, but is limited by 
the statistical power of weak 
lensing measurements. 
These are currently 
significantly more noisy than those of galaxies,
since on large scales the weak lensing signal is weak and 
the survey areas probed so far are small.
Current data sets have 
not yet reached the scales where linear theory is valid 
\citep{2002ApJ...577..604H,2003astro.ph.12036S}. In addition, 
shear calibration and background galaxy redshift distribution errors remain 
and may lead to errors as large as 20\% on the linear amplitude \cite{2004astro.ph..3255H}.
Finally, so far these studies have averaged over galaxies covering 
a broad range of luminosity over which the bias changes significantly, 
in which case comparing galaxy auto-correlation with galaxy-dark matter 
cross-correlation overestimates the bias even on linear scales, since 
$\langle b^2 \rangle /\langle b \rangle>\langle b \rangle$ (where 
$\langle \rangle$ denotes averaging over luminosity). 

Given the difficulties of the methods described above,
most workers adopt the conservative approach and 
ignore the overall amplitude of fluctuations (i.e., marginalize
over a completely unconstrained bias factor when deriving constraints
on other parameters). 
Alternatively one can use  
other methods to determine the amplitude, such as the cluster abundance or
weak lensing.
These methods have systematics of their own,
and at present various estimates of $\sigma_8$ vary by up to 30-40\% 
 (see \cite{2003astro.ph.10723T} for a recent
overview of current results). 
It would be useful to improve upon this situation, as
the overall amplitude of fluctuations 
at zero redshift is an important source of information. It is 
especially useful to constrain the dark energy equation of state 
and neutrino mass. 

The purpose of this paper is to propose a new approach to 
determine the bias parameter and to apply 
it to SDSS data. 
The main element of the method is to relate the observed luminosity 
bias to ab initio predictions of the 
halo-to-mass bias relation. 
The relation between bias and halo mass 
is one of the fundamental predictions of large scale structure models 
\citep{1989MNRAS.237.1127C,1996MNRAS.282..347M,1999MNRAS.308..119S,2001MNRAS.323....1S}.
The bias predictions depend on the cosmological model,
in particular on the nonlinear halo mass (defined below) where the bias 
is unity. Above this mass 
the bias is rapidly increasing with halo mass. 
Below this mass the bias is slowly decreasing to 
a value $b\sim 0.7$, independent of the 
other cosmological parameters \citep{1998ApJ...503L...9J,2001MNRAS.323....1S,2004astro.ph..3698S}.
If one determines observationally the bias as a function of 
halo mass one can compare it to theoretical predictions to establish
the viability of this model. 
This is particularly simple 
if one observes directly the low mass plateau, since one can
then determine the absolute bias even without accurate halo mass 
determinations. 

If we determine the correlations of dark matter 
halos as a function of their mass we also determine the overall amplitude
of fluctuations. However,
we observe galaxies, not dark matter halos, so we need to relate the two.
While it is generally accepted that all galaxies form in halos 
we also know that the relation between the two is not one to one and
galaxies of the same 
luminosity can be found in halos of different masses. For example, a 
typical galaxy like our Milky Way 
may be found at the center of a low 
mass halo with a typical size of 200kpc, it may be part of a small 
group with typical size of 500kpc or it 
may be a satellite in a cluster with a typical size of 
1-2Mpc. If we want to predict bias for a given luminosity subsample 
we must therefore determine the probability $\Delta P$
for a galaxy in this sample to be in a halo of mass $M \pm \Delta M/2$. 
To describe this we will use the conditional halo mass  probability
distribution $dP/dM \equiv p(M;{\rm L})$ 
(\citep{2002MNRAS.335..311G}, hereafter GS02).
It is important that the full distribution $p(M;{\rm L})$
is determined and not just the mean halo mass at a given luminosity. This is
because bias is a strong function of halo mass and some fraction of galaxies 
are known to be in very massive halos, which 
have a significantly larger bias than field galaxies. Even if only 
a small fraction of galaxies at a given luminosity 
are in clusters they can have a significant effect on the mean bias.

In this work we 
determine the halo mass probability distribution
for a given subsample using the 
weak lensing signal around these galaxies, the so called galaxy-galaxy 
(g-g) lensing. 
Gravitational lensing leads to
tangential shear distortions of background galaxies 
around foreground galaxies \citep{1984ApJ...281L..59T,1996ApJ...466..623B,1998ApJ...503..531H,2000AJ....120.1198F,2001ApJ...551..643S}. 
The individual distortions are small, but by 
averaging over all galaxies within a given subsample we obtain high signal 
to noise in the shear as a function of angular separation from the galaxy. 
Since we know the galaxy redshift (the foreground galaxies are taken 
from the same spectroscopic sample also used to determine the 
galaxy power spectrum), the shear signal
can be related to the projected mass 
distribution as a function of proper distance from the galaxy 
\citep{2001astro.ph..8013M}.
This allows us to determine statistically the dark matter distribution 
around any given galaxy sample. 
With g-g lensing one can determine the full halo mass function, 
since small halos contribute only at small scales, while large halos such 
as clusters give rise to a signal 
also at larger 500-2000kpc scales
typical of groups and clusters \citep{2002MNRAS.335..311G}. 
Because g-g lensing measures the signal over a wide range of scales, this allows 
one to determine the full halo mass function for a given subsample.
In this paper we will take advantage 
of the latest SDSS data compilation based on 3800 square degrees of 
spectroscopic data and imaging, a significant increase over the previous analyses 
of g-g lensing in SDSS \citep{2001astro.ph..8013M,2004AJ....127.2544S}. 

One of the key advantages of this method is that bias is a weak 
function of the halo mass, which in turn is determined with 
high accuracy from the g-g lensing analysis. For example, the ``re-Gaussianization'' method of PSF correction used for this work 
methods allows a 2-3\%
calibration error due to the PSF dilution correction \cite{2003MNRAS.343..459H}; the inclusion of other sources of systematic error as in \cite{2005astro.ph..1201M} raises the systematic error to roughly 10\% (1-$\sigma$), which leads roughly 
to a 15\% error in the 
halo mass determination. This in turn 
changes the 
bias only by 1\% around $b=1$. The effect is even smaller for masses 
well below the nonlinear mass, where the bias approaches a constant independent 
of the halo mass. This is very different from the other methods of bias 
determination discussed 
above, where the bias is (at best) a linear function of the 
signal. For example, the same 10\% weak lensing calibration error
leads to a 10\% error in bias if the large scale weak lensing-galaxy 
cross-spectrum is used as a method to determine the bias. 
The bias normalization can be predicted at any luminosity 
where the halo mass function can be determined. Since we can 
determine the halo mass function at several luminosity bins this
provides many consistency checks on the method and different bins 
can be averaged to reduce the statistical error on the bias. 

This method has other advantages as well. 
One is that we can use the same 
galaxies in the lensing analysis as in the galaxy clustering analysis. 
Using the SDSS data we can perform the g-g lensing analysis 
on the same luminosity subsamples as the ones used to obtain the large 
scale bias 
as a function of luminosity. There is no ambiguity regarding the 
selection of the catalog, which often
causes galaxies selected in 
different surveys to have different properties. Here we 
work exclusively with SDSS data.
Another advantage is that the expression 
for large scale bias weights the galaxies linearly, just as does g-g lensing.  
This is important if the number of galaxies in a halo is stochastic, 
in which case for example a pair weighted statistic (such as galaxy clustering 
on small scales) differs from a linearly weighted statistic.

G-g lensing is one of the most direct and model independent methods to determine
the halo probability distribution, short of observing the dark matter 
halos directly. In particular, this 
method avoids the problems with optical (or X-ray, SZ etc.) identification
of halos, which is only reliable for large halos such as clusters. 
Most of the galaxies are in low mass halos 
($10^{11}M_{\sun}-10^{13}M_{\sun}$), so it is important
that the mass probability distribution is reliable in that range.
Another approach to parametrize the halo occupation 
distribution is with a  
conditional luminosity function \citep{2003MNRAS.339.1057Y}. 
This was used to determine the bias in 2dF galaxies \cite{2003MNRAS.345..923V}, but in the 
absence of lensing information a more model dependent analysis
had to be used. Initial
results seemed to give higher bias than the bispectrum analysis 
of bias in 2dF \citep{2002MNRAS.335..432V}, but this may be 
a consequence of using theoretical predictions from \cite{1999MNRAS.308..119S}, which 
overestimate bias by up to 20\% compared to simulations \cite{2004astro.ph..3698S}. 


The remainder of the paper is organized as follows. In \S 2 we
present an overview of the theory, first discussing the theoretical 
predictions for halo bias and then the relation between halos and 
galaxies within the context of halo models. The analysis of the galaxy 
clustering data, weak lensing and bias is presented in \S 3.
The cosmological implications of the 
results are presented in \S 4, followed by conclusions in \S 5.  

\section{Overview of the theory \label{theory}}

In this section we first review the concept of halo bias and 
then the
formalism which relates galaxies to dark matter 
halos. We also discuss weak lensing as a method to connect the two. 

\subsection{Halo bias}

In current cosmological models structure grows hierarchically 
from small, initially Gaussian fluctuations. Once the 
fluctuations go nonlinear they collapse into virialized halos. 
The spatial density of halos 
as a function of their mass $M$ is specified
by the halo mass function $dn / dM$, which in general is a function of 
redshift $z$. It can be written as
\begin{equation}
{dn \over dM} dM={\bar{\rho} \over M}f(\nu)d\nu,
\end{equation}
where $\bar{\rho}$ is the mean matter density of the universe. We
introduced the
function $f(\nu)$, which can be
expressed in units in which it has a theoretically universal 
form independent of the power spectrum or redshift if written
as a function of peak height
\begin{equation}
\nu=[\delta_c(z)/\sigma(M)]^2.
\label{nu}
\end{equation}
Here $\delta_c$ is the linear overdensity at which a spherical perturbation
collapses at $z$ ($\delta_c=1.68$ for the spherical collapse model) and
$\sigma(M)$ is the rms fluctuation in spheres that contain on average
mass $M$ at an initial time, extrapolated using linear theory to $z$.

The first analytic model for the mass function has been 
proposed by \cite{1974ApJ...187..425P}. 
While it correctly predicts the abundance of
massive halos, it overpredicts the abundance of halos around the 
nonlinear mass scale $M_{\rm nl}$ (defined below). 
An improved version has been proposed by \cite{1999MNRAS.308..119S}.
It has been shown that 
it can be 
derived analytically within the framework of the ellipsoidal collapse model
\citep{2001MNRAS.323....1S}. 
The halo mass is defined in terms of 
the linking length parameter of the friends-of-friends (FoF) algorithm, 
which is 0.2 for the simulations used in \cite{1999MNRAS.308..119S}. 
This roughly corresponds to spherical overdensity halos of 180 times the 
background density \citep{2001MNRAS.321..372J}. For the range of
masses of interest here it is 30\% larger
than the mass defined as the mass within the radius where the 
density is 200 times the 
critical density \cite{2001A&A...367...27W,2002MNRAS.337..774S}. 
To be specific, we will use $\Omega_m=0.3$ when computing the 
virial masses, so they are defined as the mass within the 
radius within which the mean 
density is 54 times the critical density. 

Just like the underlying dark matter the halos are correlated
among themselves. The correlation amplitude depends on the halo mass, 
with more massive halos being more strongly clustered. This is 
called halo bias and can be easily understood within the 
the peak-background split of the spherical collapse model 
\citep{1989MNRAS.237.1127C,1996MNRAS.282..347M}: 
an underlying long wavelength density perturbation contributes to the 
threshold collapse value $\delta_c=1.68$, leading to a larger number 
of halos collapsing in a local overdensity of the background relative to an
underdensity.  
The more massive halos are more strongly clustered, with the strength related to 
the derivative of the mass function.

What does the halo bias depend on? While the theoretical 
predictions of \cite{1999MNRAS.308..119S} 
depend on the amplitude and shape of the power 
spectrum and the density parameter $\Omega_m$, for 
the relevant models most of the dependence can be expressed 
in terms of the nonlinear mass $M_{\rm nl}$, defined as the mass
enclosed in a sphere of radius within which the rms fluctuation 
amplitude is 1.68.  
In \cite{2004astro.ph..3698S} it was shown that a good fit to the bias relation from 
simulations is given by
\begin{eqnarray}
b(x&=&M/M_{\rm nl})=0.53+0.39x^{0.42}+{0.08 \over 40x+1}
\nonumber \\
&+&10^{-4}x^{1.7}+
\log_{10}(x)[0.4(\Omega_m-0.3+n_s-1) 
\nonumber \\
&+&0.3(\sigma_8-0.9+h-0.7)+0.8\alpha_s].
\end{eqnarray}
Here $\Omega_m$ is matter density, $\sigma_8$ is matter amplitude of 
fluctuations in spheres of 8$h^{-1}$Mpc, 
$h$ is the Hubble parameter in units of 100km/s/Mpc,
$n_s$ is the scalar slope at $k=0.05$/Mpc and 
$\alpha_s=dn_s/d\ln k$ is the running of the slope, which we approximate
as constant.
This expression should be accurate to about 0.03
over the range $0.01<x<10$.
It improves upon previous fits to simulations 
\cite{1998ApJ...503L...9J,1999MNRAS.308..119S}, 
in particularly in the regime below 
the nonlinear mass, where previous expressions overestimate the 
bias by as much as 20\%. It is clear that the dominant parameter 
for bias determination is halo mass in units of nonlinear mass, 
while the variations of cosmological parameters produce 
only small deviations from the universality of this expression. 

\subsection{Halo-galaxy connection}

In all of the current models of structure formation galaxies form inside 
dark matter halos. A galaxy of a given luminosity $L$ can form in 
halos of different mass $M$. This is described by  
the conditional halo mass  probability 
distribution at a given luminosity $p(M;L)$,
normalized to unity when integrated over mass.
The linear bias on large scales at a given luminosity is given by 
\begin{equation}
b(L)=\int  p(M;L)b(M)dM .
\end{equation}
A given cosmological model determines $b(M)$; 
to determine $b(L)$ 
we therefore need $p(M;L)$.

As mentioned in the introduction the most direct route to the 
conditional mass probability distribution
$p(M;L)$ is via g-g lensing. This measures the tangential shear 
distortions in the shapes of background galaxies 
induced by the mass distribution around foreground galaxies.
The shear distortions $\gamma_T$ are very small, in our case $10^{-3}$, 
while the typical galaxy shape noise is 0.3. To extract the signal we
must average over many foreground-background pairs. 
This results in a measurement of the shear-galaxy cross-correlation as a 
function of their relative separation on the sky. 
If the redshift of the foreground galaxy 
is known then one can express the relative 
separation in terms of transverse physical scale $R$. 
If, in addition, the
redshift distribution of the background galaxies, or their actual 
redshifts, are known, then one can relate the shear distortion $\gamma_T$ to
$\Delta\Sigma(R)=\bar{\Sigma}(R)-\Sigma(R)$, 
where $\Sigma(R)$ is the 
surface mass density at the transverse separation $R$ and $\bar{\Sigma}(R)$ 
its mean within $R$, via
\be
\gamma_T={\Delta\Sigma(R) \over \Sigma_{\rm crit}}.
\ee
Here 
\be
\Sigma_{\rm crit}={c^2 \over 4\pi G} {r_S \over (1+z_L)r_L r_{LS}},
\label{Sigma}
\end{equation}
where $r_L$ and $r_S$ are the comoving distances to the
lens and source, respectively, and $r_{LS}$ is the comoving
distance between the two (we work with comoving units throughout the paper). 
If only the probability distribution for 
source redshifts is known then this expression needs to be integrated 
over it. In principle the relation between 
angular diameter distance and measured redshift depends on cosmology, 
but since we are dealing with low redshift objects
varying cosmology within the allowed range makes little difference.
We will assume a cosmology with $\Omega_m=0.3$ and 
$\Omega_{\Lambda}=0.7$. 

We will now overview the formalism of GS02, 
beginning the discussion with a simplified description. 
Let us assume that a given halo of mass $M$ produces 
an average lensing profile $\Delta\Sigma(R,M)$. This can be obtained from a 
line of sight integration over the
dark matter profile, which in this paper is modeled as an NFW profile 
\citep{1996ApJ...462..563N} 
\begin{equation}
\rho(r)={\rho_s \over (r/r_s)(1+r/r_s)^{2}}.
\label{rho}
\end{equation}
This model assumes that the profile shape is
universal in units of scale radius $r_s$, while its characteristic density
$\rho_s$ at $r_s$ or concentration $c_{\rm dm}=r_v/r_s$ may depend on the halo mass,
which here will be modelled as $c_{\rm dm}=
10(M/M_{\rm nl})^{-0.13}$ \citep{2001MNRAS.321..559B,2001ApJ...554..114E}. 
We will define the virial radius as the radius within which the density 
is 180 times the mean density of the universe. Note that this definition 
depends on $\Omega_m$: we will adopt $\Omega_m=0.3$.
Since most of the signal is at $R>50-100h^{-1}{\rm kpc}$, baryonic 
effects can be neglected, dark matter profiles are well 
determined from simulations and concentration or the choice of the 
halo profile does not play a major role. 
The average g-g lensing signal for 
a galaxy with luminosity $L$ is 
\begin{equation}
\langle \Delta\Sigma \rangle (R;L)=\int p(M;L)\Delta\Sigma(R,M)dM. 
\label{dsp}
\end{equation}
From above we see that the same conditional mass probability 
distribution $p(M;L)$ enters in both the 
lensing signal and in the expression for bias. 
One measures the function $\langle \Delta\Sigma \rangle (R;L)$;
since the profile for individual halos is known one can invert 
the relation in equation \ref{dsp} to obtain $p(M;L)$.

Given the noisy measurements of the g-g lensing signal we cannot invert the 
conditional mass probability 
distribution with arbitrary precision, so we must assume some 
functional form for it and then fit for its parameters. 
We wish to model the probability distribution $p(M;L)$ 
in as model independent way as possible.
We will 
begin with the simplest physically motivated model and then 
add more parameters to see how the results change. By 
relaxing the assumed functional form we can test the robustness of
the final results on the model assumptions. 

The simplified description so far ignores the fact that there are 
two distinct galaxy types that need to be modeled separately. The first
type are the galaxies that formed at the centers of dark matter halos, such 
as the so called field 
galaxies or CDs sitting at the cluster centers. The second type are the non-central 
galaxies, such as satellites of Milky Way type halos or 
group and cluster members. We know that a galaxy of a given 
luminosity can be of either type, so
we split $p(M;L)$ into two parts, $p^{\rm C}$ and $p^{\rm NC}$, 
representing respectively central and non-central galaxies, 
with the fraction of non-central galaxies in each luminosity 
bin $L_i$ given by a free parameter $\alpha_i$, i.e.,
\begin{equation}
p(M;L_i) = (1-\alpha_i)~p^{\rm C}(M;L_i)+\alpha_i~p^{\rm NC}(M;L_i)~.
\end{equation}

For the central galaxy population 
we assume that 
the relation between the halo mass 
and galaxy luminosity is tight and we model this 
component with a delta-function,
\begin{equation}
p^{\rm C}(M;L_i)dM = \delta^D(M-M_{0,i})dM~,
\end{equation}
where $M_{0,i}$ are 6 more free parameters 
(we will be working with 6 luminosity bins).
In reality this component should have some width both because 
of intrinsic scatter in the $M-L$ relation and because 
we work with luminosity bins of finite width. We will ignore
this, since explicit tests have shown that the results are only weakly affected
even if the scatter is more than a factor of two in mass \citep{2002MNRAS.335..311G}. Instead, we will use simulations to account for any such effects. 

The non-central galaxies are different in that they have presumably formed 
in smaller halos which then merged into larger ones. It is thus reasonable
to assume that their luminosity
is not related to the final halo mass. Instead we 
assume a relation between the number of these non-central galaxies
and the halo mass:
the larger the halo the more satellites of a given luminosity 
one expects to find in it. We assume
this relation is a power law, $\langle N \rangle (M;L) \propto M^{\epsilon}$,
above some minimal halo mass $M_{\rm min}$, 
which should be larger than the halo mass of the central galaxy 
component above, since we are assuming that there is already another
galaxy at the halo
center.  Below this cutoff the number of 
galaxies quickly goes to zero. 
These assumptions imply
\begin{equation}
p^{\rm NC}(M;L_i) dM \propto F(M) 
M^{\epsilon_i} \frac{dn}{dM}dM~.
\end{equation}
In GS02 we have chosen $F(M)= \Theta^H(M-M_{{\rm min},i})$ where 
$M_{{\rm min},i}=3 M_{0,i}$, while here we will use a slightly more 
realistic functional form where $\epsilon=2$ below $M_{{\rm min},i}$.
We have verified that the two expressions do not differ significantly 
in final results. 
Semi-analytic models of galaxy 
formation \citep{1999MNRAS.303..188K,2002MNRAS.335..311G},
subhalos in N-body simulations \citep{2003astro.ph..8519K} as 
well as explicit comparisons with simulations \cite{2004astro.ph.10711M} 
agree with this model and predict
that for most galaxies $\epsilon \approx 1$ and $\alpha \sim 0.2$. 

For the non-central component the 
weak lensing profile $\Delta\Sigma(R,M)$
is a convolution of the halo profile with 
the radial distribution of the galaxies, which we assume to be 
proportional to the dark matter profile, $c_g=ac_{\rm dm}$. 
Observationally there is 
not much evidence for any departures from $a=1$ and we can 
test it using lensing data itself. 
Since we are explicitly excluding 
the central galaxies the non-central galaxy component of the g-g lensing 
signal does not peak 
at the center. Instead, for a given halo mass, it is small at small 
radii, peaks at a fraction of virial radius and then drops off at large radii. 
A given halo mass peaks at a given scale, so by measuring the signal over 
a broad range of scales one can extract the relative contributions of
different halo masses (note that the signal of lensing 
from neighboring halos can be neglected for $R<2$Mpc). 
See GS02 for a more detailed discussion of the
predictions of our model for lensing.

The remaining uncertainty is how much of the
dark matter around 
non-central galaxies remains attached to them. Since their 
fraction $\alpha$ is typically low ($\alpha<0.2$)
the correction due to this is small and is limited
to the inner region with $R<200h^{-1}{\rm kpc}$.
We assume the dark matter was 
tidally stripped in the outer parts of the halo, but remains
unmodified in the inner parts of the satellite halo. Effectively 
this means that each non-central galaxy also has a central contribution, 
which we model in the same way as for the central galaxies (i.e.,
as a halo with mass $M_{0,i}$ before stripping) out to $0.4r_{\rm vir}$
and totally stripped beyond that, in which case 
$\Delta \Sigma \propto R^{-2}$. This cutoff is equivalent to having 50\%
of mass stripped and agrees with simulation results in next section.

With this parametrization the mean bias in a given luminosity bin is
\be
b(L_i)=(1-\alpha_i)b(M_{0,i})+\alpha_i{\int_{M_{{\rm min},i}}^{\infty}M^{\epsilon_i}b(M)
{dn \over dM}dM \over \int_{M_{{\rm min},i}}^{\infty}M^{\epsilon_i} {dn \over dM}dM}.
\label{bl}
\ee
In the simplest form the parametrization for the conditional halo 
distribution function only has two parameters at each luminosity, 
$M_{0,i}$ and $\alpha_i$, 
with the other parameters fixed to their expected values.
Since we will be working with 6 luminosity bins this
implies a 12 parameter parametrization, which is already significantly 
more than in previous analysis of this type \cite{2001astro.ph..8013M}. Even with the order 
of magnitude increase in the data size
not all of them can be determined 
with high statistical significance. 
We will begin with these 12 parameter fits and 
then allow $\epsilon_i$ and $c_g$ to vary to see its effect on the 
final result. 
Our goal is to make the model description as  
non-parametric as possible and to show that 
our conclusions are robust against different 
parameterizations.

The halo model of GS02 is phenomenological and needs to be verified and 
possibly calibrated on simulations. A detailed comparison has 
been presented elsewhere \cite{2004astro.ph.10711M}; here we simply highlight the 
results that are of most relevance for the present study. Overall, 
the halo model is able to extract the relevant information from 
the simulations remarkably well. We find that the satellite fraction 
is determined to better than 10\%. The 
simulations reproduce well the $N(M)$ parametrization and 
indicate $\epsilon=1$. We also find that if simulations have little 
or no scatter in the mass luminosity relation then the halo model is 
able to extract the halo mass to within 10\%. However, if there is 
a significant scatter then our assumption that the central mass 
distribution is a delta function breaks down and 
there is no unique definition of halo mass. In the case of a severe
scatter there may be a significant difference between median and mean 
mass and weak lensing analysis determines something in between the two
\cite{2004astro.ph..4168T}. 
For the purpose of bias we wish to determine the mean mass or, 
even better, the full halo mass distribution. Thus the halo mass 
determined by the lensing analysis has to be increased. Here we apply 
corrections as derived in \cite{2004astro.ph.10711M} by direct comparison to 
numerical simulations. At the faint end these corrections 
are small, while for the brightest bin we apply up to a 50\% increase. 
These corrections are somewhat uncertain since we do not know the exact 
amount of scatter, so we also add a gaussian scatter with rms 0.5 of 
the correction factor to the masses from
the bootstrap resamplings to account for the additional 
uncertainty due to the scatter in mass luminosity relation. 
We emphasize again that even a 50\% correction in mass has only a small 
effect on halo bias below the nonlinear mass. 

\section{Data Analysis \label{analysis}}

We wish to determine the lensing-constrained prediction for bias as a function of 
luminosity and compare it to the observations. 
In this section we present the required procedure to achieve this goal.
This involves four steps: 

1. Determine the galaxy bias for each luminosity bin
from the galaxy clustering analysis. This step was already done in \cite{2003astro.ph.10725T}. 

2. Determine the conditional halo mass 
probability distribution from the weak lensing analysis for the 
same luminosity bins (i.e., determine the allowed values for
$\alpha_i$, $M_{0,i}$, and possibly $\epsilon_i$).  

3. Compute the predicted biases and their 
associated errors separately for each luminosity bin 
by varying over all possible configurations of 
the conditional halo mass
probability distribution consistent with the data. 
 
4. Compare the observed bias and the predictions to place constraints on the 
cosmological model. 

\subsection{Galaxy clustering analysis}

The Sloan Digital Sky Survey \citep{2000AJ....120.1579Y} uses a drift-scanning 
imaging camera \citep{1998AJ....116.3040G} and a 640 fiber double spectrograph on a
dedicated 2.5m 
telescope. It is an ongoing survey to image 10,000 sq. deg. of the sky in the
SDSS $ugriz$ AB magnitude system
\citep{1996AJ....111.1748F,2002AJ....123..485S} and to obtain
spectra for $\sim 10^6$ galaxies and $\sim 10^5$ quasars. The astrometric calibration 
is good to better than $0\farcs 1$ rms per coordinate \citep{2003AJ....125.1559P}, 
and the photometric calibration is accurate to 3\% or better 
\citep{2001AJ....122.2129H,2002AJ....123.2121S}. The data sample used for the clustering 
analysis
was compiled in Summer 2002 and is all part of 
data releases two \citep{2004astro.ph..3325A}. 
This sample consists of 205{,}443 galaxies.
For our purpose the data are divided into 6 
luminosity bins specified in table 1. 

Galaxies are selected for spectroscopic observations using 
the algorithm described in \cite{2002AJ....124.1810S}.
To a good approximation, the main galaxy sample consists of
all galaxies with $r$-band apparent Petrosian magnitude $r<17.77$.
These targets are assigned to spectroscopic plates by an adaptive
tiling algorithm \citep{2003AJ....125.2276B}.
The spectroscopic data reduction and redshift determination are performed by
automated pipelines.

\def\MM{M_{\band{0.1}r}}

\begin{table*}
\bigskip
\noindent
{\footnotesize {\bf Table 1} -- The table summarizes the luminosity
subsamples used in our analysis, listing evolution and k corrected
absolute magnitude $M_{\band{0.1}r}$ (for $h=1$), 
mean redshift $z$ in the lensing sample, 
number of foreground galaxies  used in lensing analysis, 
observed bias relative to $\MM=-20.8$ and its error, 
theoretically predicted 
bias for 2-parameter models $b_{\rm 2p}$ and its error $\sigma_{b_{\rm 2p}}$
and theoretically predicted
bias for 3-parameter models $b_{\rm 3p}$ and its error $\sigma_{b_{\rm 3p}}$.
Both fits are 
for a model with $M_0=5.6\times 10^{12}h^{-1}M_{\sun}$, which corresponds 
to $\sigma_8=0.9$, $\Omega_m=0.27$ model. For other 
values see figure \ref{fig3a}.
$M_{\band{0.1}r}$ was computed from magnitude $r$ and redshift 
$z$ assuming a flat cosmological model with
$\Omega_{\Lambda}=0.7$. Apparent magnitude cuts are $14.5<r<17.77$. 
\bigskip
\begin{center}
{\footnotesize
\begin{tabular}{cccccccccc}
\hline
Sample name     &Abs.~mag       &Mean redshift           &\# of galaxies $\,$&  $b/b_*\, $ 
& $\sigma_{b/b_*}\,$ & $b_{\rm 2p}\,$ & $\sigma_{b_{\rm 2p}}\,$ 
& $b_{\rm 3p}\,$ & $\sigma_{b_{\rm 3p}}\,$\\
\hline
{\tt L1}        &$-18<\MM<-17$  &$0.023$           & 4{,}912 & 0.723 & 0.073 & 0.67 & 0.04 & 0.67 & 0.04\\ 
{\tt L2}        &$-19<\MM<-18$  &$0.035$           & 15{,}920 & 0.764 & 0.123 &0.77 & 0.05 & 0.77 & 0.05 \\ 
{\tt L3}        &$-20<\MM<-19$  &$0.072$           &49{,}505&  0.873 & 0.077 & 0.82 & 0.03 & 0.83 & 0.03\\ 
{\tt L4}        &$-21<\MM<-20$  &$0.107$          &88{,}405& 0.969 & 0.054 & 0.85 & 0.03 & 0.85 & 0.03\\ 
{\tt L5}        &$-22<\MM<-21$  &$0.151$           &55{,}440& 1.106 & 0.063 & 1.04 & 0.05 & 1.05 & 0.05 \\ 
{\tt L6}        &$-23<\MM<-22$  &$0.205$          & 6{,}000& 1.631 & 0.119 & 1.94 & 0.20 & 1.92 & 0.25\\ 

\end{tabular}
}
\end{center}
}
\label{table1}
\end{table*}

The power spectrum analysis is described in detail in \cite{2003astro.ph.10725T}. It involves several steps. 
In the first step the group finding algorithm identifies all groups and clusters, 
which are then isotropized to remove finger of god effects. In the next step 
linear decomposition into 4000 KL modes is performed, which maximize the signal on 
large scales. These are then used in the 
quadratic estimation of the power spectra. The redshift space distortions 
(modeled using linear theory) are 
analyzed in terms of their velocity power spectrum, which is estimated together with
the galaxy power spectrum and the cross-correlation between the two. 
This analysis is performed for each of the 6 luminosity bins.

The result are 6 power spectra with similar shapes, but offset amplitudes. 
To quantify this similarity of shapes, one fits
each of the measured power spectra to the reference
$\Lambda$CDM curve with the
amplitude freely adjustable.
All six cases produce acceptable fits with reduced $\chi^2$ of order unity, and the
corresponding best-fit normalizations and associated errors are given in 
table 1 and 
shown in figure \ref{fig3b}. They are normalized relative to $L_*$ galaxies with 
$\MM=-20.8$ and are expressed in terms of linear amplitude of fluctuations 
$\sigma_8$ in figure \ref{fig3b}.

\subsection{Weak lensing analysis}

In this section we briefly review the weak lensing analysis. 
More details 
are given in a separate publication, \cite{2005astro.ph..1201M}. 
The basic model has already been outlined above. 
For each luminosity bin we parametrize the conditional halo mass
probability distribution $p(M;L)$ with a few free parameters that
we determine from the observed galaxy-lensing correlation function.
We compute this correlation function in several radial bins and use 
random sample catalogs and bootstrap resampling of real galaxies to 
determine the covariance matrix of these bins. 
We use the same 
6 luminosity bins as in the clustering analysis.
We use the SDSS sample compiled in Summer 2003 for this analysis, which consists
of $279{,}616$ galaxies, somewhat larger than what was used in the clustering 
analysis, but with identical selection criteria. 

For the background galaxies we use two samples for which the 
shape information has been extracted from the images 
using the re-Gaussianization method of \cite{2003MNRAS.343..459H},
with the implementation described in \cite{2005astro.ph..1201M}.
We require this shape information to be available in 
at least two colors. In the first sample are those with
assigned photometric redshifts, obtained by {\sc kphotoz} v3.2 \cite{2003AJ....125.2348B},
which are used for 
brighter galaxies with $r<21$. In second sample are the fainter galaxies
with $21<r<22$, for which only the expected redshift distribution is 
known.  
Details of the photometric redshift error distributions for the
brighter source sample and the
redshift distributions for the fainter sample are given in
\cite{2005astro.ph..1201M}.  The typical redshift distribution of the background sample is 
$0.1<z<0.6$.

\begin{figure*}
\centerline{\psfig{file=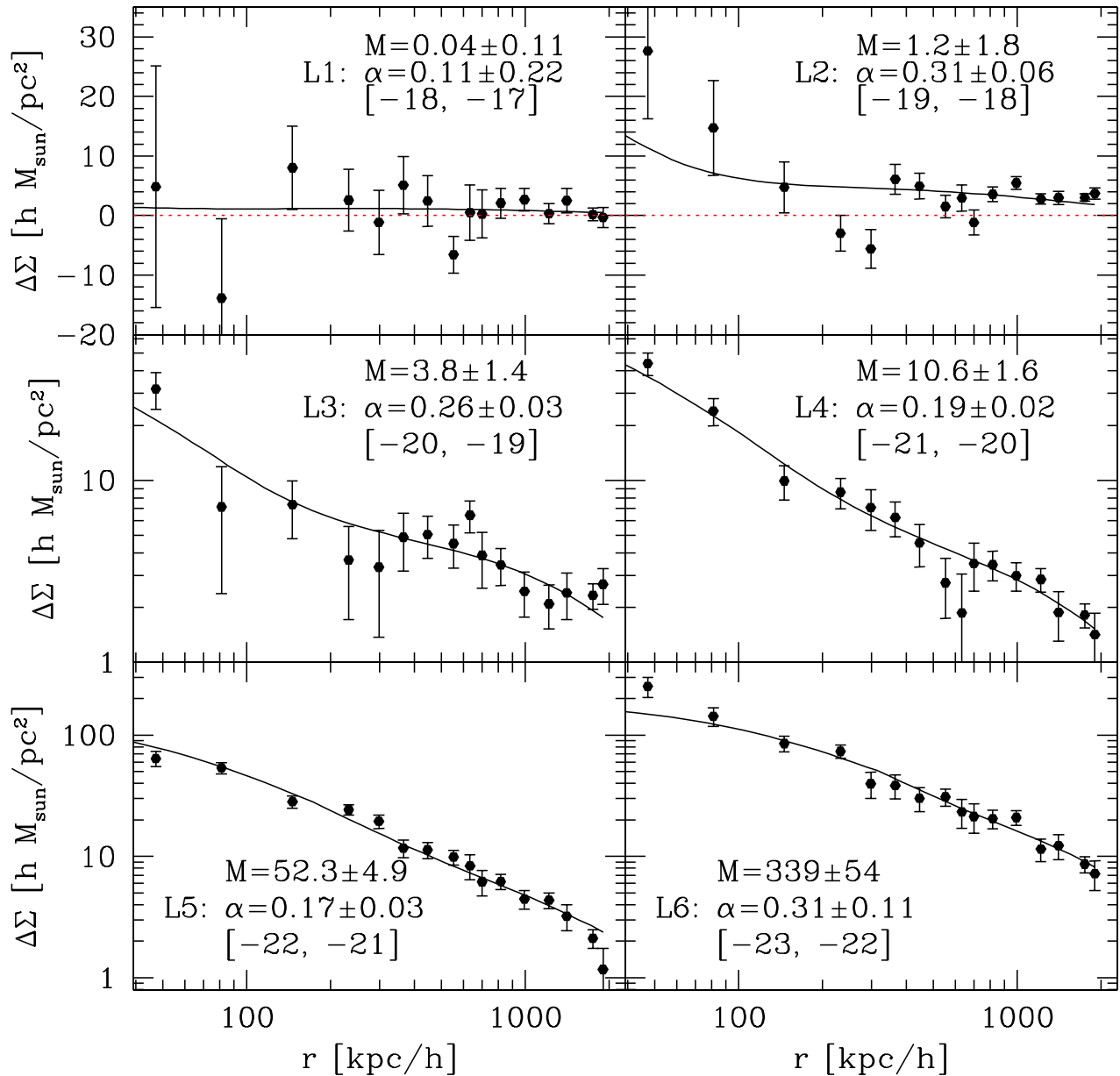,width=7in}}
\caption{Weak lensing signal $\Delta \Sigma (r)$ as a function 
of transverse separation $r$ as measured from 
SDSS data, together with the best fit 2-parameter 
model for each of 6 luminosity bins. 
Also shown are the best fit values for halo virial mass $M$ (in units of 
$10^{11}h^{-1}M_{\sun}$) and $\alpha$, the fraction of galaxies 
that are non-central, assuming $M_{\rm nl}=8\times 10^{12}M_{\sun}$. 
The model fits the data well in all bins. The mass fits are what 
comes from the fitting procedure and are a typical halo mass somewhere
between mean and median. For bias calculations they
are increased by varying 
amounts to account for the difference between fitted mass and mean 
mass, as described in the text.
}
\label{fig2}
\end{figure*}

We use a minimization routine to determine the model parameters.
A detailed description of the model and its reliability 
when applied to simulated data is presented in \cite{2004astro.ph.10711M}, 
here we just show the results in figure \ref{fig2} for
the two parameter fits, virial halo mass satellite fraction $\alpha$. 
Since we only estimate two parameters at a time 
the minimization always converges to the global minimum. 
We see that the model is 
an adequate description of the data, which is confirmed by the 
$\chi^2$ values (for L1 to L6, the $\chi^2$ values are 47, 74, 36, 36,
51, and 44 respectively for 44 degrees of freedom). Note that for all
but the faintest bin  
there is a clear detection of the 
signal and both the central 
and noncentral components are determined.
In some of the brightest bins the signal to noise in these detections 
is enormous compared to previous analyses of this sort \cite{2001astro.ph..8013M}. 
The virial mass of the central component 
scales nearly linearly with luminosity in 
all but the brightest bin (there is no detection in the faintest bin), 
while the non-central fraction 
is roughly constant at $\alpha=0.13$ except in the faintest bin where 
it is much lower and in the brightest bin where it is much higher. 
These are the values assuming $M_{\rm nl}=8\times 10^{12}M_{\sun}$. 
This value increases as the nonlinear mass decreases, since the mass 
function is exponentially cut-off above nonlinear mass, so the 
abundance of high mass halos is reduced and to fit the observed signal 
the fraction $\alpha$ must increase. We do the fits 
on a grid of values for $M_{\rm nl}$ that spans the range of interest. 
In the 
faintest bin no detection of the noncentral component is obtained 
and the central component is marginal. However, the 
resulting upper limits are still useful, since they imply that a
majority of these galaxies cannot live in massive halos,
otherwise we would have detected a stronger lensing signal.
For low mass halos the bias is only weakly dependent on mass,
so even an upper limit leads to a strong 
constraint on the bias. In fact, if the low mass plateau where $b\sim 0.65-0.7$ 
could be reached  then one could determine the absolute
bias directly just from this low luminosity population without 
any additional modeling. In practice whether or not 
this  low mass limit is
reached with the range of halo masses we can probe here depends on the 
value of nonlinear mass, 
but the fact that the bias is flattening at the faint end does 
place useful constraints when compared to observations, where the 
same trend is observed. 

We repeat the analysis by adding more parameters to the fit.
The most relevant parameter for bias 
is $\epsilon$, which changes the relative proportion of 
less massive versus more massive halos for the non-central component. 
We find that we cannot determine all 3 parameters separately at each 
luminosity bin, so that $\epsilon$ is strongly anti-correlated with $\alpha$. 
For example, 
we find that even solutions with $\epsilon=2$ are allowed, suggesting 
a large fraction of massive halos. However, at the same time $\alpha$
is reduced, so that the overall fraction of massive halos is more or
less unchanged. 
This is exactly what one would expect, since the signal at $R=300-2000h^{-1}$kpc
is measuring directly the contribution from these 
massive halos (groups and clusters). In other words, the 3-parameter fit
is essentially a non-parametric fit to the data with large degeneracies
between the parameters, but little variation in the halo mass 
probability distribution in the relevant range.

We also tried a different nonparametric fit, where we divide 
the halo masses into several bins and fit for the fraction in each separately. 
Not surprisingly 
we find huge degeneracies between individual components 
in this case, especially at the low mass end. Even in this 
most general case the 
fraction of galaxies residing in large halos (above $10^{14}h^{-1}M_{\sun}$) 
is constrained to be below 0.2 at the bright end and below 0.01 at the faint 
end. It is particularly important that the fraction of galaxies in 
these high mass halos is strongly constrained, 
as these halos have the largest bias and a poor constraint on them
leads to a larger error on the bias predictions. 
The variations of bias consistent 
with these various halo mass probability distributions are in fact very small.

While for standard analysis we assumed that radial distribution of 
galaxies is the same as that of dark matter, we also explored other 
possibilities. Weak lensing data have some sensitivity to determine 
the radial distribution directly, since a shallow galaxy 
distribution also leads to a shallow radial dependence of $\Delta \Sigma$
\cite{2004astro.ph.10711M}. However, this only makes a difference at large 
radii and the central mass determination only weakly depends on it. We 
find that the central halo mass changes by less than 10\% if $c_g$ is 
varied by a factor of 2 from its assumed value $c_{\rm dm}$. 

One possible source of systematical error is 
the shear calibration from measured ellipticities.
For this analysis using re-Gaussianization, as shown in
\cite{2005astro.ph..1201M}, when we include all sources of shear calibration
error (not just the 2-3\% PSF dilution correction), we are left with a
roughtly 10\% shear calibration error ($2\sigma$).

The redshifts of background galaxies are another possible source
of systematic error. 
For those 
galaxies that have
photometric redshifts the main difficulty is knowing the error 
distribution, since even a relatively minor fraction of outliers can 
skew the distribution and lead to a bias in the lensing signal. 
This is particularly problematic for the brighter galaxies ($r<21$)
for which the typical redshift is 0.2-0.4. 
We use direct matching of SDSS objects 
with deeper spectroscopic surveys (DEEP2) to calibrate our photometric
redshifts.  
An independent  analysis using Luminous Red 
Galaxies (LRG) for which we know the redshift error distribution, 
provides an independent confirmation of our photoz calibration. 
Details of the LRG redshift distributions are given in
\cite{2004astro.ph..7594P}, and 
the comparison with the other methods used for this paper is in
\cite{2005astro.ph..1201M}. 

The redshift distributions as a function of magnitude are relatively 
well measured at these magnitudes, 
so one could use those instead of photometric 
redshifts. However,
the existing redshift distributions that apply to the overall 
population of galaxies  
cannot be directly used in the lensing analysis, especially 
not at the faint end. This is because a 
significant fraction of galaxies are rejected or downweighted
in the lensing analysis because they are too small to give reliable
ellipticity measurements. These galaxies tend to be smaller and 
thus at a higher redshift relative to the overall population. 
So the effective redshift of the ``lensing weighted'' population tends 
to be lower than that of the overall population at the same 
magnitude. The same effect may 
also lead to an underestimate of $\sigma_8$ in weak lensing 
measurements of the power spectrum, 
where it can have much more damaging consequences. 
It is very difficult to account for this effect if one 
does not have the complete redshift information of a representative 
portion of the data. 

Besides the shear calibration and redshift distribution, there are
several other sources of systematic error in the weak lensing signal:
magnification bias, stellar contamination, intrinsic alignments, sky
level determination errors.  We use estimates of the values of these
errors for our analysis from \cite{2005astro.ph..1201M}, and place a 10\%
$1-\sigma$ overall calibration error on the lensing signal on top of
the (comparable) statistical errors.

\subsection{Bias predictions}

In the next step we take the results from the previous section 
to compute the predicted bias as a function of luminosity. 
Bias is a nonlinear function of the model parameters and we 
wish to determine both the mean and the variance. 
The fits for the halo mass probability distribution can be 
strongly degenerate, so a Gaussian approximation for the fitted 
parameters is not necessarily valid. In addition, one must impose
physical constraints such as $\alpha>0$ and $M_0>0$. 
Our approach is to use bootstrap resampling to determine the 
errors. We divide the observed area into 200 chunks of roughly 
equal area. We then bootstrap resample these, by randomly choosing 
200 chunks with replacement (so 
some of these are duplicated). 
We perform the fitting 
procedure 
as described above on each of the bootstrap realizations. We use
the fitted parameters to compute 
the bias using equation \ref{bl}.
Finally, we compute the mean and variance of the bias parameter by 
averaging over 2500 of these bootstrap resamplings. 

Before discussing the results we need to address the redshift evolution. 
Redshift evolution can affect the weak lensing analysis, the bias analysis 
and the clustering analysis. 
Most of the galaxies are at low redshift up to $z\sim 0.2$ 
(table 1),
so any corrections due 
to the redshift evolution are small. Moreover, there are various competing 
effects that further suppress the redshift evolution effects. 

We work in comoving coordinates, so redshift effects are minor. 
One usually applies the evolution 
correction to the galaxy luminosity to account for the fact that 
higher redshift galaxies are brighter because their stars are younger.
We use the correction as given in \cite{2003ApJ...592..819B}, adding
the quantity $1.6(z-0.1)$ to the $r$-band absolute magnitudes. 
This correction is not important as long as we use the same definition 
of luminosity  in our weak lensing analysis as is in the 
clustering analysis.
The definition of the virial mass as defined
by FOF algorithm of simulations is in comoving coordinates: 
FOF algorithm finds clusters with a density contrast 
of order $b^{-3}$ relative to the mean, 
where $b$ is the linking length of FOF. This 
definition does not vary with redshift in comoving coordinates. 
The main effect of redshift evolution is that nonlinear mass at 
higher redshifts is lower, which affects the mass function and 
bias predictions. We perform the  
analysis on a grid of nonlinear masses defined at z=0 and for each
use the nonlinear mass value appropriate for the median redshift for 
a given luminosity bin.
Regarding the clustering evolution with redshift, 
for $L_*$ galaxies a typical redshift is 0.1, so 
once we find bias for these galaxies and multiply it with the observed 
amplitude of galaxy clustering we need to evolve this 
amplitude of fluctuations to 
redshift 0, which increases it by roughly 5\% in a $\Lambda$CDM
universe.  

\begin{figure}
\centerline{\psfig{file=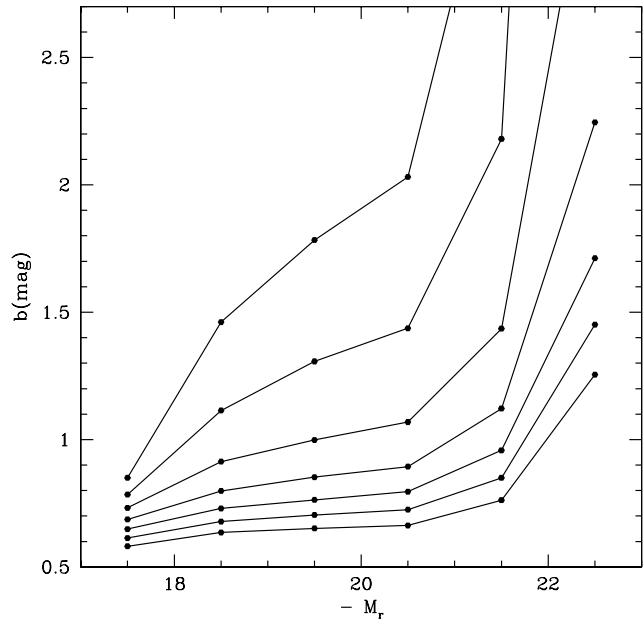,width=3.5in}}
\caption{
This figure shows the lensing-constrained model predictions for bias as a 
function of nonlinear mass using the 2-parameter 
models of the halo mass probability distribution. More general models
of the halo probability distribution
give very similar results and are not shown here. The nonlinear masses 
from top to bottom are 
$3.4 \times 10^{11}h^{-1}M_{\sun}$,
$6.2 \times 10^{11}h^{-1}M_{\sun}$,
$1.7 \times 10^{12}h^{-1}M_{\sun}$,
$4.0 \times 10^{12}h^{-1}M_{\sun}$,
$8.0 \times 10^{12}h^{-1}M_{\sun}$,
$1.5 \times 10^{13}h^{-1}M_{\sun}$ and
$2.4 \times 10^{13}h^{-1}M_{\sun}$.
Errors have been suppressed (see Table 1). 
}
\label{fig3a}
\end{figure}

The results for 2-parameter models
are shown in figure \ref{fig3a} for several values of 
the nonlinear mass $M_{\rm nl}$, spanning the range 
from $1.6\times 10^{11}h^{-1}M_{\sun}$ 
(corresponding to $\sigma_8=0.6$, $\Omega_m=0.2$ model) at the top 
to $2.5\times 10^{13}h^{-1}M_{\sun}$ 
(corresponding to $\sigma_8=1.1$, $\Omega_m=0.3$ model) at the bottom 
(we vary $\sigma_8$ from 1.1 to 0.6 from bottom up assuming $\Omega_m=0.3$, 
the top one has $\sigma_8=0.6$ and $\Omega_m=0.25$).
Note that the weak lensing determination of $\alpha$ depends on 
$M_{\rm nl}$, since the cluster contribution depends on the cluster 
mass function: for higher value of nonlinear mass the exponential 
cutoff in the mass function is at a higher mass and so the fraction 
of galaxies in this component can be lower to match the observational 
constraints. We include this by performing the lensing 
analysis on all the values of $M_{\rm nl}$ of interest and use that 
information when computing bias predictions as a function of $M_{\rm nl}$. 

One can see that as a consequence of 
the weak lensing determination of the halo mass  distribution models with 
low nonlinear masses (low $\sigma_8$ and/or $\Omega_m$) 
predict higher bias than those with higher nonlinear masses. 
Notice how the models with high nonlinear mass predict almost constant 
bias with luminosity, a consequence of bias being independent of mass 
below 0.1$M_{\rm nl}$. On the other hand, if nonlinear mass is close 
to the halo mass of $L_*$ galaxies then the bias is rapidly changing 
with luminosity. 

We can address the robustness of the bias predictions by comparing 
the results between 2 and 3-parameter models. 
These are shown in table 1 for nonlinear mass 
$M_{\rm nl}=5.6\times 10^{12}h^{-1}M_{\sun}$, corresponding to the
$\sigma_8=0.9$, $\Omega_m=0.27$ model. 3-parameter models are 
very degenerate and often give unlikely values 
such as $\epsilon=0$ (corresponding to the case where the number 
of galaxies within a halo is constant regardless of halo mass). 
This is compensated by increasing the value of $\alpha$ so that 
the data are still fit well. Remarkably, the mean bias and its error 
hardly change from 2-parameter models to 3-parameter models in all 
bins. 
This demonstrates that the bias predictions are robust against the 
parametrization of the halo probability distribution.

\begin{figure}
\centerline{\psfig{file=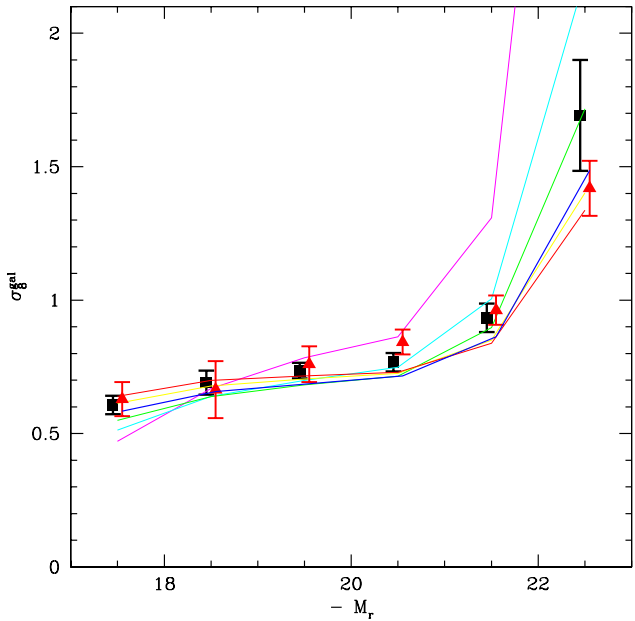,width=3.5in}}
\caption{
The halo bias predictions of galaxy fluctuation amplitude $\sigma_8^{\rm gal}$ as a 
function of luminosity varying linear matter amplitude $\sigma_8$: 
0.6, 0.7, 0.8, 0.9, 1.0, 1.1,
from top to bottom
on the right hand side. 
The remaining parameters have been 
fixed to $\Omega_m=0.3$ and $n_s=1$. 
Squares are for model with $\sigma_8=0.88$, $n_s=1.0$ and 
$\Omega_m=0.3$. For this model we show errors from theoretical modelling.  
Also shown as triangles are the observed values of $\sigma_8^{\rm gal}$. 
}
\label{fig3b}
\end{figure}

Figure \ref{fig3b} shows the predicted values for galaxy clustering 
amplitude $\sigma_8$ for models 
with $\sigma_8$ varying from 0.6 to 1.1, with $\sigma_8=0.88$ in bold. 
This is obtained by 
taking predicted bias values and multiply them with $\sigma_8$. 
Also shown is the observed values of $\sigma_8$
which are obtained by taking $b/b_*$ values in table 1 and 
multiplying them with $\sigma_{8}(\MM=-20.8)=0.875$ value as obtained 
from \cite{2003astro.ph.10725T}. The latter is almost independent on cosmological parameters 
and has a statistical error of 0.013.
The first thing to notice is the  agreement 
between the theoretical predictions and the observations, with 
theoretical predictions being slightly lower than observations at L4. 
Theoretical models predict both the gradual flattening of bias for galaxies 
fainter than L4, as well as a rapid increase in bias from L5 to L6, both 
of which are observed in SDSS data. 
This rapid increase in bias from L5 to L6 is caused both by the rapid
increase in bias above the nonlinear mass as well as by the rapid decrease in 
star formation efficiency for the most massive halos: going from 
L5 to L6 we increase the luminosity by a factor of 2.5, while the halo mass has increased
by a factor of 6.5 (figure \ref{fig2}), and the fraction of non-central galaxies
has also increased. 
So the scaling between halo mass and luminosity becomes much steeper 
at the bright end and most of the galaxies in the $[-22,-23]$ bin reside
in group and cluster halos with masses above $10^{13}h^{-1}M_{\sun}$.

This agreement between the theoretical bias predictions and observations 
suggests that 
a fundamental prediction of large scale structure models, that of 
the bias dependence on halo mass,
has been confirmed. 
Very low values of $M_{\rm nl}$
predict that bias is rapidly 
changing with mass over the observed range, 
which is not observed. 
On the other side of the mass range, 
very high values of $M_{\rm nl}$ predict bias 
is changing very slowly with mass, which would also be in contradiction 
with the observations. 
These results constrain $\sigma_8$, but these constraints by themselves 
are not very strong. 
In fact, one 
can find good agreement 
for $\sigma_8\sim 0.7-1.0$. 

At first this agreement between theoretical predictions and 
observations is so good it is almost disappointing, 
since the two agree over a broad range of $M_{\rm nl}$. 
While this result confirms the 
basic prediction of the structure formation models it would appear
that it does not allow us to 
place useful constraints on cosmological models. 
This conclusion is too pessimistic, since nonlinear mass is also 
changed by varying $\Omega_m$ and slope $n_s$. For example, reducing 
$\Omega_m$ to 0.27 from 0.3 and $n_s$ to 0.96 from 1 reduces nonlinear 
mass by a factor of 2 and increases theoretical bias predictions by 
10\% at L4. This brings the observations into a better agreement  
with theoretical predictions and improves the fit to $\chi^2 \sim 7$  
at $\sigma_8=0.88$.
The data thus favor slightly lower values of $\Omega_m$ or $n_s$ than their 
canonical values of 0.3 and 1, respectively. 
We note that this analysis is strongly sensitive on
the accuracy of the bias as a function 
halo mass relation and we would find very different conclusions 
using expressions in \cite{1998ApJ...503L...9J,1999MNRAS.308..119S} over the more recent ones in \cite{2004astro.ph..3698S}. 

These heuristic arguments are formalized in the next section, where we 
incorporate the bias constraints into the parameter estimation procedure.
It is clear from this discussion that the bias constraints 
depend on nonlinear mass, which in turn depends on several cosmological 
parameters such as $\sigma_8$, $\Omega_m$ and $n_s$, 
so the best strategy is to perform a full analysis over 
the parameter space of interest. 

However, it is worth highlighting where the strongest constraints are 
coming from and explore if the bias determination of flux limited 
sample would improve the cosmological constraints. 
One can see from figure \ref{fig3a}
that for $L4$ in the range $\sigma_8=0.6-0.9$ all models give essentially 
the same value of $\sigma_8(L4)\sim 0.75$, 
which is lower than the observed value 
of 0.85.  
Similarly, we can take the predictions for bias as a function of 
luminosity and weight it by galaxy numbers (given in table 1)
to obtain the bias prediction for the flux limited sample. This 
has the advantage that it can be related to the observed value 
which has very small errors, $\sigma_8(\MM=-20.8)=0.88\pm 0.02$. 
However, the predictions are almost independent of $\sigma_8$: 
we find 0.77 at $\sigma_8=0.7$ and 0.73 at $\sigma_8=1$ for 
$\Omega_m=0.3$ and $n_s=1$. The fact that the predictions are 
lower than observed value again argues that these two 
parameters (or shape parameter $\Gamma$ which depends on Hubble 
parameter $h$ as well) must be lower. 
Therefore, using the flux limited sample amplitude  
one {\it cannot} 
determine the bias and $\sigma_8$ 
despite a relatively small error on the 
observationally determined amplitude of galaxy clustering. The reason 
for that is the degeneracy in the way bias changes with amplitude of 
fluctuations: reducing $\sigma_8$ by 10\% reduces $M_{\rm nl}$ by a 
factor of 2 and increases bias predictions around $M=M_{\rm nl}$ by 10\%, 
so the product $b\sigma_8$, which determines the galaxy clustering 
amplitude, remains unchanged. We find that
by using the full range of luminosity breaks this degeneracy, 
because the bias is very slowly changing with nonlinear mass on one end 
and very strongly changing with it on the other end of luminosity range. 

\subsection{Bias error budget}

We will include the 
following $\chi^2$ component in the overall likelihood evaluation,
\be
\chi^2 = \sum_{i=1}^{6}{\left( b_{\rm model,i} - b_*(b/b_*)_{\rm i}\right)^2
\over \sigma_{b_{\rm model,i}}^2+b_*^2\sigma_{b/b_*,i}^2+\sigma_{\rm sys}^2},
\label{chi2}
\ee
where $b_{\rm model,i}$ is the predicted bias for the i-th luminosity bin and 
$\sigma_{b_{\rm model,i}}$ the corresponding error,  
$(b/b_*)_{\rm i}$ is the observed bias at the same luminosity and 
$\sigma_{b/b_*,i}$ the corresponding error and
$\sigma_{\rm sys}=0.03$ accounts  
for systematic uncertainties in the theoretical
modeling of the bias and its variations with the model \cite{2004astro.ph..3698S}. 
For a given model we first compute $M_{\rm nl}$ and then interpolate
between the values shown in figure \ref{fig3a} to obtain $b_{\rm model,i}$ and 
$\sigma_{b_{\rm model,i}}$. Note that $b_*$ is one of 
the parameters we are varying and is 
constrained both by the $\chi^2$ above 
and by the overall amplitude of the galaxy power spectrum.
For $b_{\rm model,i}$ we use the 2-parameter fits as given in table 
1, although using the 3-parameter fits would give almost 
identical results.

Equation \ref{chi2} contains 3 contributions to the bias error. First 
is the error on the theoretical bias prediction, 
$\sigma_{b_{\rm model,i}}$. This error is dominated by the uncertainties
in the conditional halo mass probability distribution $p(M;L)$. 
Despite significant uncertainties in the probability distributions
(specially for the 18 parameter fits) the resulting values of 
$\sigma_{b_{\rm model,i}}$ are between 
0.02-0.08 for $M_{\rm nl}\sim 10^{13}h^{-1} M_{\sun}$. 
This is typically smaller than the errors on the observed 
bias $\sigma_{b/b_*,i}$, which range from 0.054 to 0.12. 
We assign a systematic uncertainty   
$\sigma_{\rm sys}=0.03$ to all of the bins except the brightest one (L6),
where we use $\sigma_{\rm sys}=0.1$ to account for larger variations 
in model predictions, 
as well as larger systematic uncertainties 
due to the rapid variation of the bias with luminosity and redshift. 
The systematic error is 
subdominant compared to the clustering or lensing error. 
Current bias
constraints are mostly dominated by the observational uncertainties
in the bias from the clustering analysis and 
not by the modeling uncertainties of either the bias or conditional 
halo mass probability distribution as determined by weak lensing.
Note that systematic uncertanties in calibration and redshift 
distribution have already been included in the lensing analysis and 
corresponds to about 0.03 in bias. 
The current statistical error in the bias, averaging over all 6 bins in table 
1, is 0.03, so the systematic error is at most equal to 
the statistical error. 

Much of the error budget in $\sigma_{b/b_*,i}$ is due to the sampling variance. 
We are treating the bias estimates as uncorrelated between luminosity bins, 
assuming they come from 
independent volumes. In reality 
there is some overlap in volume between neighboring luminosity bins 
and some of the same large scale structure contributes to two bins
at the same time (see e.g. figure 3 in \cite{2003astro.ph.10725T}). 
Because of this our error estimation is conservative, 
since for overlapping regions sampling variance between luminosity 
subsamples should be reduced. The reduction is however rather modest 
even for very large scales \cite{2004astro.ph..3698S}. 
Note that systematic uncertainties lead to some correlations between the 
errors, which we ignore in the present analysis since they are small. It
would be straightforward to generalize upon this by computing the 
correlations between the bootstrap samples. 

\section{Cosmological parameter determination \label{cosmo}}

In this section we include the bias constraints in the parameter determination
procedure to see if we can constrain cosmological models better than 
without this information. 
We combine the constraints from the SDSS power spectrum 
with CMB 
observations from WMAP \citep{2003ApJS..148....1B,2003ApJS..148..135H,2003ApJS..148..161K}. 
We implement the Monte Carlo Markov Chain method \citep{2001PhRvD..64b2001C} using 
CMBFAST version 4.5\footnote{available at cmbfast.org} \cite{1996ApJ...469..437S}, 
outputting
 both the CMB spectra and the corresponding matter power spectra $P(k)$.
We evolve all the matter power spectra to a high $k$ using CMBFAST and 
we do not employ any analytical approximations. In addition, we use 
linear to nonlinear mapping of the matter power spectrum using expressions
given in \cite{2003MNRAS.341.1311S}. 

Our implementation of the MCMC is the same as in \cite{2003MNRAS.342L..79S}. It 
is independent of that used in \cite{2003astro.ph.10723T}, but we
verified that the 
results for the case of WMAP+SDSS without bias agree. 
A typical run is based on 48 independent chains, 
contains 50,000-200,000 chain elements and requires 2-4 days 
of running on a 48 processor cluster in a serial mode of CMBFAST\footnote{
While CMBFAST is parallelized with MPI we 
found that running it in parallel results in about a factor of 2 penalty on 
8 processors (and more if more processors are used), 
mostly due to the fact that the highest $k$-modes take the longest to run. 
The current implementation distributes
$k$-modes to individual processors, 
so the 
master node must wait for the slowest $k$ mode to finish before the final 
assembly.
For 48 processors the additional premium
due to the required burn-in of each chain does not offset this penalty, 
so one is better off running CMBFAST serially. If significantly more 
processors were used the cost of burn-in would increase and 
one would be better off running CMBFAST 
in parallel in 8 node batches.}. 
The success rate was of order 30-50\%,
correlation length (as defined in \cite{2003astro.ph.10723T}) 10-30
and the effective chain length of order 3,000-20,000. 
We use 23-39 chains and
in terms of Gelman and Rubin $\hat{R}$-statistics \citep{gelman92} we find the
chains are sufficiently converged and mixed, with $\hat{R}<1.05$,
ie we are more conservative than the recommended value $\hat{R}<1.2$.

Our pivot point is at $k=0.05/$Mpc and we use the
tensor normalization convention in which for the simplest 
inflationary models the tensor to scalar ratio is 
$r=T/S=-8n_T$. Our most general parameter space is
\be
\bi{p}=(\tau,\omega_b,\omega_m,m_{\nu},\Omega_{\lambda}, {\cal R}, n_s,
\alpha_s,T/S,b_*),
\ee
where $\tau$ is the optical depth, $\omega_b=\Omega_bh^2$ is proportional
to the baryon 
to photon density ratio, $\omega_m=\Omega_mh^2$ is proportional to the matter
to photon density ratio, $m_{\nu}$ 
is the massive neutrino mass, 
$\Omega_{\lambda}$ is the dark energy density today and $w$ its equation 
of state, ${\cal R}$ is the amplitude of curvature perturbations at 
$k=0.05$/Mpc, $n_s$ is the scalar slope at the same pivot and 
$\alpha_s=dn_s/d\ln k$ is the running of the slope, which we approximate 
as constant. We fix the tensor 
slope $n_T$ using $T/S=-8n_T$. 
We do not allow for non-flat
models, since curvature is 
already tightly constrained by CMB and 
other constraints,
which leads to $\Omega_{K}=0.02\pm 0.02$ 
for the simplest models \cite{2003ApJS..148..175S}. 
For the more general models, such as those with dark energy 
equation of state, relaxing this assumption can lead to a significant 
expansion of errors. We are therefore testing a particular class of 
models with $K=0$ 
and not presenting model independent constraints on equation of state.
We follow the WMAP team in imposing a $\tau<0.3$ constraint. Upcoming 
polarization data from WMAP will allow a verification of this prior. 
From this basic set of parameters we can obtain constraints on 
several other parameters, such as the baryon and matter densities 
$\Omega_b$ and $\Omega_m$, Hubble parameter $h$ and amplitude of 
fluctuations $\sigma_8$. 
Since we do now allow for curvature one has
$\Omega_{\lambda}=1-\Omega_m$ and we use $\Omega_m$ in 
table 2. In fact, our primary parameter 
is $\Theta_s$, the angular scale of the acoustic horizon, which is tightly 
constrained by the CMB. 
Similarly, although we use ${\cal R}$ as the primary parameter in 
the MCMC we present the amplitude in terms of the more familiar $\sigma_8$. 

The basic result for two different MCMC runs are given in table
2 for SDSS combined with WMAP.
For most of the parameters we quote the median value (50\%), 
[15.84\%,84.16\%] interval ($\pm 1\sigma$), and [2.3\%,97.7\%] interval 
($\pm 2\sigma$). These are calculated from the cumulative one-point 
distributions of MCMC values for each parameter and do not depend on the 
Gaussian assumption. For the parameters without a detection we only quote 
a 95\% confidence limit. All of the restricted 
parameter space fits are acceptable based on $\chi^2$ values, starting 
from the 6-parameter model with no tensors, running or neutrino mass. 
Introducing additional parameters does not improve the fit significantly.
However, we wish to determine the amplitude of fluctuations in an 
as model independent way as possible and for this reason we explore the 
most general parameter space possible. 

Below we discuss the results from this table in more detail. 
Our standard model has 6 cosmological parameters: $\tau,\omega_b,\omega_m,\Omega_{\lambda}=1-\Omega_m, {\cal R}, n_s$, plus ``nuisance parameter'' bias $b_*$. 
For the case without bias our 
results are in a broad agreement with those in \cite{2003astro.ph.10723T}, 
although a slightly different treatment (modification of lowest
multipoles in WMAP and inclusion
of $\tau<0.3$ constraint) does lead to small changes in the 
best fit parameters and their errors. 

\begin{figure}
\centerline{\psfig{file=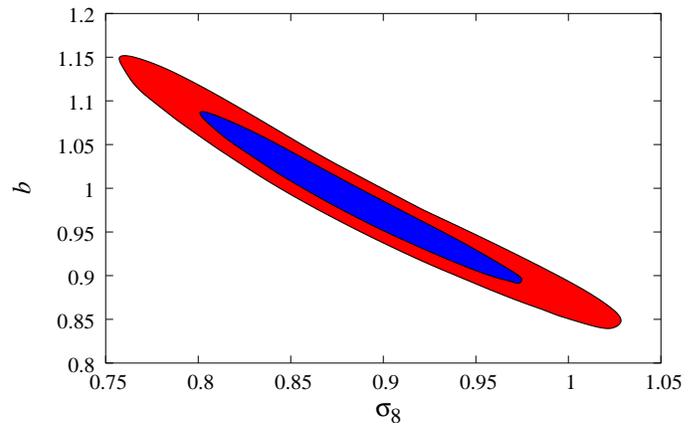,width=3.5in}}
\caption{68\% (inner, blue) and 95\% (outer, red) 
contours in $(\sigma_8,b_*)$ plane
using SDSS+WMAP+bias measurements. 
The two parameters are strongly correlated because only their 
product is determined from SDSS clustering analysis. The additional
bias constraint helps reduce the degeneracy. 
}
\label{fig4}
\end{figure}

\subsection{$\sigma_8$ and bias}

\begin{table*}
\noindent
{\footnotesize
Table 2: median value, 
$1\sigma$ and $2\sigma$ 
constraints on cosmological parameters combining CMB, SDSS power spectrum 
shape and SDSS bias information.
The columns compare different theoretical priors.
The parameters for 7 parameter models are
$\tau,\omega_b,\omega_m,\Omega
_{\lambda}=1-\Omega_m, {\cal R}, n_s$, plus ``nuisance parameter'' bias $b_*$.
\begin{center}
\begin{tabular}{|l|c|c|c|c|c|}
\hline 
& 7par    & 7par+T/S &7par+$m_{\nu} $ & 7par+$\alpha_s$ & 7par+T/S+$m_{\nu}$+$\alpha_s$ 
\\
& & & & &   \\
\hline
$10^3\omega_b$ & 
$23.4^{+1.2}_{-1.1}{}^{+2.5}_{-2.1}$ &
$25.1^{+1.6}_{-1.5}{}^{+3.4}_{-2.8}$ &
$23.7^{+1.3}_{-1.2}{}^{+2.5}_{-2.2}$ &
$22.9^{+1.4}_{-1.4}{}^{+2.8}_{-2.7}$ &
$24.8^{+1.6}_{-1.6}{}^{+3.3}_{-3.1}$ 
\\
& & & & &   \\
$\Omega_m$ & 
$0.253^{+0.027}_{-0.026}{}^{+0.053}_{-0.047}$ &
$0.226^{+0.027}_{-0.026}{}^{+0.055}_{-0.053}$ &
$0.259^{+0.031}_{-0.027}{}^{+0.074}_{-0.048}$ &
$0.269^{+0.041}_{-0.033}{}^{+0.091}_{-0.062}$ &
$0.262^{+0.051}_{-0.036}{}^{+0.138}_{-0.072}$ 
\\
& & & & &   \\
$n_s$ & 
$0.987^{+0.037}_{-0.031}{}^{+0.071}_{-0.055}$ &
$1.040^{+0.045}_{-0.042}{}^{+0.094}_{-0.078}$ &
$0.995^{+0.037}_{-0.034}{}^{+0.065}_{-0.060}$ &
$0.959^{+0.052}_{-0.053}{}^{+0.104}_{-0.106}$ &
$1.00^{+0.054}_{-0.058}{}^{+0.118}_{-0.121}$ 
\\
& & & & &   \\
$\tau$ & 
$0.181^{+0.068}_{-0.066}{}^{+0.110}_{-0.116}$ &
$0.187^{+0.063}_{-0.062}{}^{+0.103}_{-0.119}$ &
$0.202^{+0.064}_{-0.072}{}^{+0.092}_{-0.130}$ &
$0.195^{+0.065}_{-0.068}{}^{+0.097}_{-0.123}$ &
$0.232^{+0.046}_{-0.064}{}^{+0.064}_{-0.127}$ 
\\
& & & & &   \\
$b_*$ & 
$0.984^{+0.070}_{-0.065}{}^{+0.129}_{-0.119}$ &
$0.965^{+0.068}_{-0.062}{}^{+0.131}_{-0.113}$ &
$1.02^{+0.079}_{-0.074}{}^{+0.157}_{-0.137}$ &
$0.970^{+0.069}_{-0.060}{}^{+0.133}_{-0.106}$ &
$0.986^{+0.078}_{-0.065}{}^{+0.158}_{-0.115}$ 
\\
& & & & &   \\
$\sigma_8$ & 
$0.884^{+0.064}_{-0.057}{}^{+0.120}_{-0.098}$ &
$0.904^{+0.062}_{-0.60}{}^{+0.121}_{-0.105}$ &
$0.854^{+0.066}_{-0.060}{}^{+0.127}_{-0.112}$ &
$0.896^{+0.058}_{-0.058}{}^{+0.108}_{-0.104}$ &
$0.882^{+0.062}_{-0.063}{}^{+0.116}_{-0.119}$ 
\\
& & & & &   \\
$h$ & 
$0.732^{+0.034}_{-0.031}{}^{+0.065}_{-0.056}$ &
$0.773^{+0.042}_{-0.038}{}^{+0.097}_{-0.071}$ &
$0.728^{+0.034}_{-0.034}{}^{+0.067}_{-0.069}$ &
$0.716^{+0.039}_{-0.039}{}^{+0.078}_{-0.079}$ &
$0.738^{+0.045}_{-0.050}{}^{+0.100}_{-0.112}$ 
\\
& & & & &   \\
$T/S$ &   0 & $<0.49$ (95\%) & 0 & 0 
& $<0.57$ (95\%)
\\
& & & & &   \\
$m_{\nu}$ & 0 & 0 & $<0.18$eV (95\%) & 0 & $<0.24$eV (95\%)  
\\
& & & & &   \\
$\alpha_s$ &  0 & 0 & 0 & $ -0.024^{+0.031}_{-0.031}{}^{+0.062}_{-0.061} $ &
$ -0.045^{+0.036}_{-0.040}{}^{+0.073}_{-0.089}$ 
\\
& & & & &   \\

\hline
\end{tabular}
\end{center}
}
\label{table3}
\end{table*}

\begin{figure}
\centerline{\psfig{file=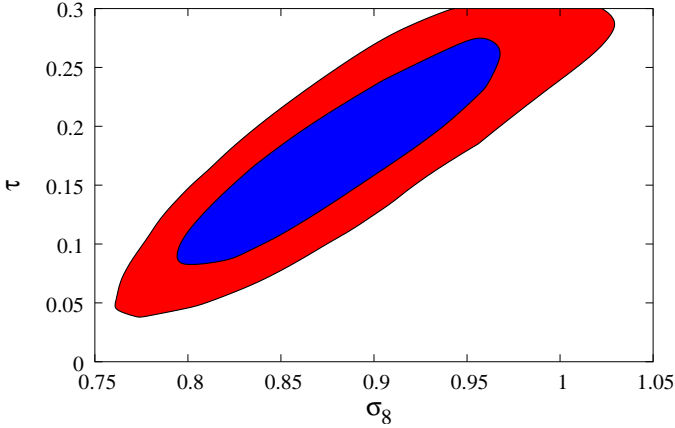,width=3.5in}}
\caption{68\% (inner, blue) and 95\% (outer, red) 
contours in $(\sigma_8,\tau)$ plane
using SDSS+WMAP+bias measurements. There is a correlation between the 
two, so a better determination of optical depth $\tau$ from 
polarization data would help improve the constraints. 
}
\label{fig5}
\end{figure}

Figure \ref{fig4} shows the 68\% and 95\% contours in the 
$(\sigma_8,b_*)$ plane.
The two parameters are strongly correlated because 
the SDSS power spectrum constrains their product to be $b_*\sigma_8=0.87\pm 0.02$. 
Both bias and $\sigma_8$ are in a good 
agreement with the SDSS+WMAP analysis {\it without} bias constraints, 
which gives  
for the basic 6-parameter model 
$b_*=0.96\pm 0.08$ and $\sigma_8=0.92\pm 0.09$ \cite{2003astro.ph.10723T}, but which in the 
presence of massive neutrinos changes to $b_*=1.06\pm 0.10$ and $\sigma_8=0.82 \pm 0.09$. 
We find 
$\sigma_8=0.88\pm 0.06$ and $b_*=0.99\pm 0.07$ 
in the presence of neutrinos and running with 
bias constraints included. 
There remains a significant
degeneracy between $\sigma_8$ and optical depth $\tau$, as shown in 
figure \ref{fig5}.
The upcoming WMAP polarization analysis may help improve
this degeneracy. 

Let us now compare these constraints to other methods to determine 
bias and $\sigma_8$. The closest analysis to ours is that of the
WMAP+2dF bispectrum using $k\sim (0.2-0.3) h$/Mpc modes.
Our $\sigma_8$ is in a good 
agreement with the WMAP+2dF analysis with the bias constraint from the bispectrum, 
which gives $\sigma_8=0.84\pm 0.04$. 
The 2dF bispectrum analysis has been 
performed on scales smaller than 
the scale of the power spectrum analysis ($k < 0.2h/$Mpc), so many of 
subsequent analysis papers chose not to adopt this constraint. 
The fact that a completely 
independent approach presented here finds the same result is therefore
encouraging and suggests that the systematics are not dominating 
the statistical errors in either approach. 

There are several other methods of $\sigma_8$ determination, such as cluster 
abundance, weak lensing and Sunyaev-Zeldovich effect.
Cluster abundance 
estimations of $\sigma_8$ range from 0.6 to 1, often with very small errors 
(see
a recent overview of the current situation in \cite{2003astro.ph.10723T}).
Many of these are reported for $\Omega_m=0.3$, so if the actual value is 
somewhat lower the required value for $\sigma_8$ increases. 
The main difficulty is in calibrating the mass-temperature relation, 
which cannot be done with simulations, because these still lack some of the 
physics of cluster formation such as cooling, feedback, conduction etc. 
Direct 
calibration from the mass and temperature measurements on individual 
clusters is more 
promising, but is limited by statistical and systematic errors.
With this method the results are particularly 
sensitive to calibration errors at various steps of the analysis, 
so the challenge for the future will be to
control them at the required level. 

Weak lensing
observations also have a similar spread in reported 
values of $\sigma_8$ 
(between 0.7 and 1, see overview in \cite{2003astro.ph.10723T}). The assigned statistical
errors are larger, so there may not be much conflict among 
different observations. In addition, with this method there are also systematic 
calibration effects at the 10-20\% level in $\sigma_8$. Some of 
these are discussed in in the context of present analysis
in \S III.B, but a similar 
discussion applies to other weak lensing analyses as well. 
These are often not included in the error budget. 
As discussed above, 
these arise both for shear calibration from ellipticity measurements and 
for the redshift distribution of background galaxies. 

Finally, CBI \cite{2003ApJ...591..540M,2004astro.ph..2359R} 
and BIMA \cite{2002ApJ...581...86D} 
measurements of the CMB at high $l$ find excess power, 
which can be interpreted as a Sunyaev-Zeldovich signal. If so this 
would require a fairly high normalization, with estimates of $\sigma_8$ 
ranging from $0.98^{+0.06}_{-0.07}$ \cite{2004astro.ph..2359R}
to 
$1.04\pm 0.12 ({\rm stat}) \pm 0.1 ({\rm sys})$ \citep{2002MNRAS.336.1256K}, 
where  
both errors are 2-sigma. Some of the
difference between the two estimates is due to the drop of the CBI
amplitude in the latest analysis, while the rest is due to the differences
in the modeling of the signal from either simulations or analytic models. 

In summary, our value of $\sigma_8$ is consistent with most recently 
reported values. It is at the lower end of what is
required to explain the CBI/BIMA excess 
power in terms of the SZ effect and at the upper end of some of the
weak lensing and cluster abundance measurements. Each one of 
these may have additional systematic errors 
that could bring the results into a better agreement. 
Our values are in excellent agreement with  
the WMAP+2dF analysis. 

\subsection{Neutrino mass} 

Galaxy surveys are important as tracers of neutrino mass, since 
neutrinos have a considerable effect on the matter power 
spectrum. At the time of decoupling such 
neutrinos are still relativistic,
but
become nonrelativistic later in the evolution of the universe if 
their mass is sufficiently high.
Neutrinos free-stream out of their 
potential wells, erasing their own perturbations on scales smaller
than the so-called free streaming length, defined as the distance at which a neutrino of a given
rms velocity $v_{\rm th}$ can still escape against gravity.
The velocity is $c$ when neutrinos are relativistic and drops as $1/a$
afterwords because of momentum
conservation, so 
\be
v_{\rm th} \sim {k_BT \over m_{\nu}}=50(1+z)(m_{\nu}/{\rm eV})^{-1}{\rm km/s}.
\label{vth}
\ee
Since the comoving Hubble time is proportional to
$\tau_H \sim [(1+z)\Omega_m]^{-1/2}H_0^{-1}$
the product of the two gives an estimate of the free-streaming length.
The resulting comoving free-streaming wavevector is (for one 
massive family)
\be
k_{fs}=0.4(\Omega_m h^2)^{1/2}(1+z)^{-1/2}{m_{\nu} \over 1{\rm eV}} {\rm Mpc}^{-
1}.
\label{kfs}
\ee
It should be evaluated at redshift when neutrinos become non-relativistic, since
the 
dominant contribution comes from when neutrinos are relativistic. 

\begin{figure}
\centerline{\psfig{file=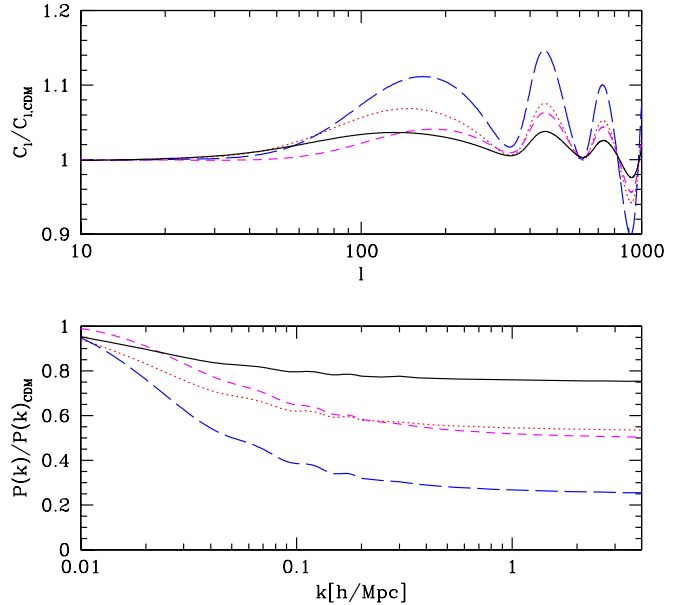,width=3.5in}}
\caption{Top panel shows the change in CMB spectrum $C_l$ for several 
neutrino masses relative to 
zero mass. The masses are $m_{\nu}=0.15$eV (solid, black), 0.3eV (dotted, red)
and 0.6eV (long dashed, blue), all 
with 3 neutrino families of equal mass. Also shown (short dashed) is the case of
3 massless + 1 massive neutrino family with $m_{\nu}=0.9$eV. Bottom shows the 
ratio of matter power spectra for the same models. We see that while 
increasing neutrino mass increases the CMB spectrum it decreases the
matter power spectrum. For the same total mass the 3+1  
model is more non-relativistic 
at recombination, has a 
smaller effect on the CMB spectrum relative to 3 families of equal mass and
the corresponding mass limits are weaker. 
}
\label{fig9}
\end{figure}

For a given wavevector $k$ neutrino perturbations
are suppressed while
$k>k_{fs}$. After that they can grow again and may even catch up with
the matter perturbations.
When neutrinos are dynamically important
the neutrino 
damping also affects the matter fluctuations, decreasing their amplitude
on scales below the free streaming length. One can see from equation \ref{kfs}
that the scale is fairly large
for the neutrino masses of interest, $k_{\rm fs}\sim 0.1{\rm Mpc}^{-1}$
at $z=0$ for $m_{\nu}  \sim 1eV$. 
Below this suppression scale the power spectrum shape is the same as in 
regular CDM models, so on small scales 
the only consequence is the suppression of the amplitude (see figure \ref{fig9}). 
We can thus adopt the halo bias 
predictions from CDM models and apply them to massive neutrino models as well. 
We should note that while qualitatively the effects are similar for 1 or
3 massive neutrino families, they differ in detail (figure \ref{fig9}), 
so the constraints 
are not directly comparable and one must do a separate analysis in 
the two cases. We mostly focus on 3 degenerate 
neutrino families here, but also present MCMC results for 3 massless+1 massive 
family below.

While it is commonly believed that massive neutrinos have a minor 
effect on the CMB, this is actually not entirely the case for the masses 
of interest below 2eV. This is because neutrinos with such a
low mass are still relativistic when they enter the horizon for 
scales around $k=0.1 h$/Mpc and are either relativistic or 
quasi-relativistic at the time of recombination, $z\sim 1100$ (equation 
\ref{vth}). As a result neutrinos cannot 
be treated as a nonrelativistic component with regard to the CMB.
Figure \ref{fig9} shows how much the CMB spectrum changes for various neutrino 
masses relative to the zero mass case, keeping $\Omega_m=\Omega_{\rm cdm}+
\Omega_b+\Omega_{\nu}$ constant (as well as the other 
cosmological parameters). One can see that for $m_{\nu}=0.3$eV
massive neutrinos increase the spectrum by
6\% at $l=200$, well above the errors (neutrinos are also not 
degenerate with respect to the CMB if we compare them to a 
fixed $\Omega_{\rm cdm}+\Omega_b$ instead). 
In addition, massive neutrinos
increase the CMB spectrum but suppress the 
power spectrum (figure \ref{fig9}), which enhances the sensitivity
when the two tracers are combined. 

We note that our bias constraints significantly improve
upon neutrino mass limits. 
In the absence of biasing constraints the neutrino mass limit 
from WMAP+SDSS is 
$m_{\nu}<0.6$eV if their masses
are nearly degenerate \cite{2003astro.ph.10723T}.
Biasing constraints
improve significantly upon this. We find
\be
m_{\nu}<0.18{\rm eV} (95\%) ~{\rm (3~families,~without~running)},
\ee
at 95\% for a
single component if we assume no running, as was
done in all of the work to date.
Our constraints improve upon WMAP+2dF
constraints, where $m_{\nu}<0.23$eV was found
by combining WMAP and 2dF with the
bias determination from the
bispectrum analysis \citep{2002MNRAS.335..432V}.

However, this result was 
based on the assumption of no running. This assumption was
present in all of the work so far, including \cite{2003ApJS..148..175S}
where it was argued that there is a weak evidence of running. 
A running spectral index 
changes the shape of the power spectrum as 
the massive neutrinos do, so including the running in the parameter 
estimation can significantly expand the limits. 
Naive expectations are that a negative running (which seems to be 
preferred by the data) suppresses the power 
on small scales just as a massive neutrino and so would lead to a 
tighter constraint on neutrino mass, but MCMC analysis 
does not confirm this and the constraints are weakened. 
With biasing constraints and running our constraint changes to
\be
m_{\nu}<0.24{\rm eV} (95\%) ~{\rm (3~families,~with~running)}.
\ee

All of the mass limits presented here are based on 3 degenerate
massive neutrino families. If one assumes a model with 3 massless
families and 1 massive family (a sterile neutrino model),
as motivated by LSND \cite{1996PhRvL..77.3082A},
then the mass limits on the sum change, since both the CMB and
and the transfer function change. One finds the
limits are significantly weakened: in the WMAP+2dF analysis without bias
the limit is 1.4eV \cite{2003JCAP...05..004H}.
We find the same
\be
m_{\nu}<1.37{\rm eV} (95\%) ~{\rm (3+1~families,~no~running)}.
\ee
The reason for the relatively weak constraint is that this case is much
more degenerate with $\Omega_m$ than the case of 3 degenerate massive
neutrino families.
From the LSND experiment the allowed regions are four islands with the lowest mass
$m_{\nu}=0.9$eV  and the next lowest 1.4eV \cite{2003NuPhS.114..203M}.
We see that the windows are
rapidly closing with cosmological constraints, but the case is not yet
air tight.

There were recent claims that the neutrino mass may have already been
detected from cosmological observations \citep{2003astro.ph..6386A}. These claimed
detections are inconsistent with our WMAP+SDSS or with WMAP+2dF
constraints and are based
primarily on the cluster abundance analysis of \cite{2003MNRAS.342..287A},
which seems to
prefer a low value for the amplitude, $\sigma_8\sim 0.7$.
As discussed above, cluster abundance estimates of $\sigma_8$ range
from 0.6 to 1 and are likely to be dominated by systematics often not
included in the quoted errors, such as mass-temperature calibration.
Increasing the error on this method to account
for systematics removes the evidence for neutrino mass.

\section{Discussion and Conclusions}

In this paper we presented a detailed comparison between observations and 
theoretical predictions of one of the fundamental predictions of structure
formation models, that of linear bias as a function of halo mass. In 
doing so we combine two separate observational analyses of SDSS data, 
the galaxy clustering analysis and a weak lensing
analysis, both as a function of galaxy luminosity.
The former gives us the relative bias as a function of luminosity, 
while the latter connects the galaxies of a 
given luminosity class to their dark matter halo mass distribution.
We find a remarkable agreement between the observations and theoretical 
predictions of bias over a range of halo masses from 
$10^{11}h^{-1}M_{\sun}$ to $10^{13}h^{-1}M_{\sun}$.
This success should be viewed as an important new confirmation of the current 
large scale structure paradigm 
in predicting the properties of the universe we live in.

The second goal of this work is to provide a determination of 
the bias for SDSS galaxies, which can be used to improve the cosmological 
parameter estimation. 
For any given model we can determine bias from the 
halo mass-bias relation and from the amplitude of galaxy clustering.
The two must agree, which requires the bias of $\MM=-20.8$ 
galaxies to be very close to unity. 
As a result we can place constraints
on the amplitude of fluctuations,
$\sigma_8  = 0.88\pm 0.06$, as well as on the 
other other cosmological parameters. 
Our results are in an excellent agreement with 
the WMAP+2dFGRS analysis of \cite{2003ApJS..148..175S}. 
In particular, we find no evidence for 
any systematic differences between the SDSS and 2dF power spectra
in either amplitude or shape. 

The systematic errors 
from the galaxy clustering data have been thoroughly examined in \cite{2003astro.ph.10725T}, 
but some open question remain to be addressed. 
One of them is the correction for 
nonlinear effects in the power spectrum analysis. 
These can affect both the conversion from
redshift space to real space and from the real space power spectrum to 
the linear power spectrum.
The current analysis in \cite{2003astro.ph.10725T}
is based on the power spectrum 
with $k<0.2 h$/Mpc, but this cutoff is somewhat arbitrary and 
should be 
justified within a more realistic model, which will provide an 
estimate of the systematic error as a function of $k$. Currently the nonlinear 
corrections are based on the nonlinear evolution model of \cite{2003MNRAS.341.1311S}. 
However, galaxies are not a perfect tracer of dark matter and
the nonlinear correction for galaxies could be different from that 
of dark matter. For example, in the context of halo models nonlinear 
effects are entirely due to the correlations within the halos. If 
galaxies populate larger (smaller) halos than the dark matter then the 
nonlinear corrections will be larger (smaller). 
The halo model is not sufficiently accurate to address 
these questions in detail and simulations are needed instead.  
To put things in perspective, the overall galaxy clustering 
amplitude $b_*\sigma_8$ from 
SDSS using $k<0.2 h$/Mpc data points has an error of 1.5\%, while the 
nonlinear correction at $k=0.2 h$/Mpc is around 10\%. 
In this situation it does not take
much for the systematic error to dominate over the statistical error. 
However, much of the error will be on the overall amplitude and this 
is still limited by the error on our bias determination, which is 
around 7\%.
Nonlinear effects are likely to be 
even more important for the luminosity dependent analysis of 
galaxy clustering, which was used in this paper as a basis for 
bias determination, but statistical errors in this analysis are 
larger and systematics may not dominate the results. It is clear 
that these issues have to be revisited if one is to believe the
cosmological implications from these results. 
 
We have argued that the method presented here is in many ways
more robust than some of the 
other methods to determine the bias from observations. 
Still, there remain possible systematic errors in the present analysis that 
need to be explored further.  
Two uncertainties mentioned in the present analysis are
the calibration of the weak lensing signal and the accuracy of the bias-halo 
mass relation. We have argued that the weak lensing method is robust 
in that even a 20\% calibration error leads to only 0.03 error in bias. 
This does not dominate relative to the statistical error and we have 
included it in the analysis. 
Similarly, the bias-halo mass relation has been calibrated to an accuracy 
of 0.03 using a suite 
of large simulations covering some of the parameter space of interest
\cite{2004astro.ph..3698S}, but larger simulations and more extensive grid of parameter 
space is needed to improve this to an accuracy of 0.01. 

The current paper should be viewed as a first application of this method 
of bias determination,
rather than the last word on it. There are many ways the current analysis
could be improved. The most important among these is reducing the error on the 
clustering amplitude as a function of luminosity, specially
for low luminosity galaxies. As we argued most of the leverage for bias 
determination comes from the low luminosity galaxies, which reside in low 
mass halos, and for which the bias is only weakly dependent on the
nonlinear mass (it is also relatively insensitive to errors in the 
lensing analysis). A better analysis optimized to reduce the sampling variance 
errors should reduce the errors considerably.
Galaxies with absolute luminosities in the range $[-18,-20]$ seem 
particularly promising, since they have a reliable lensing detection 
(figure \ref{fig2}) and a weak bias dependence on nonlinear mass (figure 
\ref{fig3a}), unless the nonlinear mass is very low. 
Their clustering amplitude is currently very
poorly determined compared to the overall sample (Table 1), which 
could be improved dramatically with a more careful analysis. 
In addition, the systematic errors in the lensing analysis could be 
reduced further.
All of these aspects can be improved in the near future. 
This could lead to significant improvements on the cosmological parameters
such as neutrino mass or equation of state. 

The present paper is only the first in several possible 
attempts to estimate the large scale bias in SDSS. An ongoing project
closest to our approach 
is to use the galaxy auto-correlation function on small scales to 
constrain the halo occupation probability distribution.  
A bispectrum analysis  
of SDSS galaxies is also in progress and should yield results which are 
statistically comparable to the present analysis. 
A weak lensing analysis on large scales can also determine
the bias, although this would require larger survey areas
than currently available
and a tight control of all possible 
systematics. Current efforts are limited to small scales and their
statistical power remains weak \cite{2003astro.ph.12036S}. 
Finally, with better modeling of redshift space distortions the 
constraints on $\beta$ may also improve beyond the current 
limits, which at the moment remain weak \cite{2003astro.ph.10725T}.
Combining and comparing these with the current analysis will 
provide additional checks of systematics in these methods. 

The method presented here can be applied to other samples of 
galaxies, such as those selected 
by color, spectral type or stellar mass.
Of particular interest would be to apply it to 
the higher redshift galaxies, such as 
the Luminous Red Galaxies, which 
are very numerous in surveys such as SDSS 
and whose photometric redshifts are relatively 
accurate and well understood. 
Their clustering amplitude on large scales
can be determined with a high 
accuracy, close to 1\%, in several redshift bins up to $z=0.7$. 
Without a model for bias the amplitude does not give useful information.
If one could determine their halo mass distribution 
function with lensing that would
allow one to predict the bias and thus extract
the growth factor as a function of redshift. While the absolute 
calibrations of bias are still difficult at a 1\% level, the
relative calibration as 
a function of redshift may be more promising. This may be one of the 
most promising methods to place constraints on the dark energy equation 
of state and its evolution. 

Funding for the creation and distribution of the SDSS Archive has been provided
by the Alfred P. Sloan Foundation, the Participating Institutions, the National
Aeronautics and Space Administration, the National Science Foundation, the U.S.
Department of Energy, the Japanese Monbukagakusho, and the Max Planck Society.
The SDSS Web site is http://www.sdss.org/. 
We thank M. Strauss, P. Steinhardt and L. Verde for useful comments. 

The SDSS is managed by the Astrophysical Research Consortium (ARC) for the
Participating Institutions.  The Participating Institutions are The University
of Chicago, Fermilab, the Institute for Advanced Study, the Japan Participation
Group, The Johns Hopkins University, Los Alamos National Laboratory, the
Max-Planck-Institute for Astronomy (MPIA), the Max-Planck-Institute for
Astrophysics (MPA), New Mexico State University, University of Pittsburgh,
Princeton University, the United States Naval Observatory, and the University
of Washington.

Our MCMC simulations were run on a Beowulf cluster at Princeton University, 
supported in part by NSF grant AST-0216105.
US is supported by fellowships from the 
David and Lucile Packard Foundation, Alfred P. Sloan Foundation,
NASA grants NAG5-1993, NAG5-11489 and NSF grant CAREER-0132953.
MT was supported by NSF grant AST-0134999, NASA grant
   NAG5-11099 and fellowships from the David and Lucile
   Packard Foundation and the Cottrell Foundation.

Funding for the DEEP2 survey has been provided by NSF grant
AST-0071048 and AST-0071198. Some of the data presented herein were
obtained at the W.M. Keck Observatory, which is operated as a
scientific partnership among the California Institute of Technology,
the University of California and the National Aeronautics and Space
Administration. The Observatory was made possible by the generous
financial support of the W.M. Keck Foundation. The DEEP2 team and Keck
Observatory acknowledge the very significant cultural role and
reverence that the summit of Mauna Kea has always had within the
indigenous Hawaiian community and appreciate the opportunity to
conduct observations from this mountain. 

 \bibliography{apjmnemonic,cosmo,cosmo_preprints}

\end{document}